\documentclass[fleqn,usenatbib]{mnras}

\usepackage{newtxtext,newtxmath}
\usepackage{xparse}
\usepackage[dvipsnames]{xcolor}

\usepackage[T1]{fontenc}

\DeclareRobustCommand{\VAN}[3]{#2}
\let\VANthebibliography\thebibliography
\def\thebibliography{\DeclareRobustCommand{\VAN}[3]{##3}\VANthebibliography}

\usepackage{graphicx}	
\usepackage{amsmath}	
\usepackage{xspace}

\DeclareMathOperator{\id}{{\rm d}\!}

\newcommand{\fornax}{F{\sc{ornax}}\xspace}

\NewDocumentCommand\pder{mmg}{\ensuremath{
		\IfNoValueTF{#3}
		{\dfrac{\partial #1}{\partial #2}}
		{\left(\dfrac{\partial #1}{\partial #2}\right)_{#3}}
}}

\defcitealias{Burrows2020}{B20}

\title[Proto-neutron Star Spins and Kicks]{Kicks and Induced Spins of Neutron Stars at Birth}

\author[M. S. B. Coleman et al.]{
Matthew S. B. Coleman,$^{1}$\thanks{E-mail: msbc@astro.princeton.edu}
Adam Burrows$^{1}$
\\
$^{1}$Department of Astrophysical Sciences, 4 Ivy Lane, Princeton University, Princeton, NJ 08544, USA
}

\date{Accepted XXX. Received YYY; in original form ZZZ}

\pubyear{\the\year{}}

\begin{document}
\label{firstpage}
\pagerange{\pageref{firstpage}--\pageref{lastpage}}
\maketitle

\begin{abstract}
Using simulations of non-rotating supernova progenitors, we explore the kicks imparted to and the spins induced in the compact objects birthed in core collapse.  We find that the recoil due to neutrino emissions can be a factor affecting core recoil, comparable to and at times larger than the corresponding kick due to matter recoil. This result would necessitate a revision of the general model of the origin of pulsar proper motions. In addition, we find that the sign of the net neutrino momentum can be opposite to the sign of the corresponding matter recoil. As a result, at times the pulsar recoil and ejecta can be in the same direction. Moreover, our results suggest that the duration of the dipole in the neutrino emissions can be shorter than the duration of the radiation of the neutron-star binding energy. This allows a larger dipole asymmetry to arise, but for a shorter time, resulting in kicks in the observed pulsar range. Furthermore, we find that the spin induced by the aspherical accretion of matter can leave the residues of collapse with spin periods comparable to those inferred for radio pulsars and that there seems to be a slight anti-correlation between the direction of the induced spin and the net kick direction.  This could explain such a correlation among observed radio pulsars. Finally, we find that the kicks imparted to black holes are due to the neutrino recoil alone, resulting in birth kicks $\le$100 km s$^{-1}$ most of the time.
\end{abstract}




\section{Introduction}

Core-collapse supernova (CCSN) explosions give birth to neutron stars and stellar-mass black holes. The former have been studied for decades as radio pulsars \citep{Manchester2005,Kaspi2016}, and more recently as magnetars \citep{1998Natur.393..235K,Kaspi2017}, while the latter have been studied predominantly as a subclass of X-ray sources \citep{Shakura,Remillard}, and more recently in the context of potential LIGO/Virgo binary gravitational wave sources \citep{Ligo}.
It is estimated that as many as 10$^8$ neutrons stars and at least 10$^6$ stellar-mass black holes created in the terminal collapse of the cores of massive stars ($\ge$8 M$_{\odot}$) reside in the Milky Way. Hence, the properties of such exotic compact objects are centrally important to an understanding of our galaxy and to interesting systems within it.

The explosion energy and nucleosynthetic isotope yields are primary features of CCSNe.  Along with the associated birth spectrum of neutron-star and black-hole masses and the morphology of the explosion debris fields, they constitute important quantities to be explained with theory and simulation \citep{Burrows2021}. However, equally important is the explanation of the origin of pulsar proper motions and spins. The birth kicks and periods constitute ground truth for any detailed model of their origin in supernovae.

The pulsar proper motions range from near zero km s$^{-1}$ to more than $\sim$1500 km s$^{-1}$. Their mean values are in the range 350$-$400 km s$^{-1}$. Importantly, these mean and extreme speeds are generally too large to be explained solely, or even mainly, by the Blaauw mechanism due to unbinding of progenitor binaries \citep{Blaauw,Hills,Blaauw2}. Therefore, neutron stars must generically be kicked during the supernova explosion itself.  Some observations hint at a bimodal kick distribution \citep{Cordes1998,Arzoumanian2002,Verbunt2017}, while others suggest a simpler distribution \citep{Faucher2006,Chatterjee2009}.  There is a suggested correlation of kicks with supernova remnant asymmetries \citep{Holland-Ashford2017, 2018ApJ...856...18K}. Recently, GAIA has provided measurements of kicks over a wide range of speeds \citep{Yang2021}.  The existence of bound neutron stars in globular clusters requires that some neutron stars are born with low kick speeds (below $\sim$50 km s$^{-1}$) \citep{Lyne1994,Arzoumanian2002}.  Finally, analysis of stellar-mass black holes in X-ray binaries suggests average kick speeds less than $\sim$80 km s$^{-1}$ \citep{Fragos2009}, with a few possible exceptions \citep{Mandel2016,Fragos2009,Repetto2017}.

Pulsar spins evolve significantly with time and measured pulsar spin periods range from milliseconds to many seconds \citep{Faucher2006,Lyne1994,Manchester2005,Popov2012,Noutsos2013}.
Most of the millisecond pulsars are likely spun up by later accretion from a companion, but they constitute an identifiable subset of radio pulsars. Using spin-down models, birth spins of the other pulsars are back-computed to range 1) from $\sim$100 to $\sim$300 milliseconds (ms) \citep{Chevalier1986}; 2) with a mean of $\sim$200 milliseconds \citep{Igoshev2013}; 3) up to several hundred milliseconds with no multimodality \citep{Faucher2006}; 4) with a Gaussian distribution with mean and variance of $\sim$100 milliseconds \citep{Popov2012}; or 5) with any broad distribution less than $\sim$500 milliseconds \citep{Gullon2014}. There is some evidence for a spin-kick correlation at birth for radio pulsars \citep{Wang2006,Wang2007,Johnston2005,Johnston2007,Ng2007,Noutsos2013}.

Theoretical models for pulsar and neutron-star birth kicks can be divided into five categories:
1) models that invoke an off-center spinning magnetic dipole and the asymmetrical radiation of the nascent relativistic radio pulsar wind \citep{Harrison1975} (however, see \citet{Lai2001}). This model is not currently favored;
2) models that rely on neutrino radiation in the context of rapid rotation \citep{Spruit1998}.  Such models are a bit simplistic, lack detailed support from sophisticated simulations, but naturally predict spin-kick correlations;
3) models requiring kicks due to asymmetrical late-time fallback \citep{Janka2021}.  How much fallback is generic is still very much unknown, and the process may be rare or associated with only a small subset of progenitors.  Nevertheless, such a model naturally produces spin-kick alignment;
4) models that rely on kicks from asymmetrical supernova matter ejecta and the associated proto-neutron star (PNS) recoil \citep{Scheck2006,Nordhaus2010,Nordhaus2012,Wongwathanarat2013,Janka2017,Nakamura2019} or solely on kicks from neutrino emission dipole asymmetries \citep{2006ApJS..163..335F, Nagakura2019b}.
The former ejecta recoil models have gained currency, and incorporate compelling aspects that have emerged from more sophisticated 2D \citep{Scheck2006} and 3D \citep{Wongwathanarat2013} simulations, while the latter addresses asymmetries that arise from more sophisticated neutrino radiative transfer approaches \citep{Nagakura2019b}; and 5) models that manifest kicks due to both asymmetrical neutrino radiation and asymmetrical matter ejection seen in sophisticated 3D supernova explosion simulations \citep{Stockinger2020,Burrows2020}. This explanation is the most compelling.  However, for the lion's share of the most sophisticated extant models, the asymptotic state is rarely reached in the associated expensive simulations. Nevertheless, this is the model we explore in this paper, and for one of these models we have indeed reached that asymptotic state.

It would seem that matter recoil kicks naturally correlate with the degree of ejecta asymmetry and the explosion energy \citep{Burrows2007,Janka2017}.  Moreover, the neutrino kick relies on the persistence of an interesting degree of neutrino emission asymmetry that persists for the many seconds of deleptonization and cooling.  The net asymmetry ($\alpha_{\nu}$) need not be large, however, as the following formula for the resultant kick ($v_{\nu}$) attests:
\begin{align}
    v_\nu\sim\alpha_{\nu} \dfrac{E_\nu}{c M_{\rm PNS}}=120\ {\rm km\ s}^{-1}\left(\dfrac{\alpha_\nu}{10^{-2}}\right)\left(\dfrac{E_\nu}{10^{53}\ {\rm ergs}}\right)\left(\dfrac{M_{\rm PNS}}{1.4M_{\sun}}\right)^{-1}\, ,
\label{nkick}
\end{align}
where $E_{\nu}$ is the total radiated neutrino energy and $M_{\rm PNS}$ is the residual proto-neutron star (PNS) mass. For a net asymmetry of only 1\%, a radiated energy of $3\times 10^{53}$ ergs translates into a kick of $\sim$360 km s$^{-1}$ \citep{Burrows1996}.  However, it is not at all clear a priori the degree to which the matter and neutrino effects add or cancel. This is one aspect we will attempt to address in this paper.

The most natural explanation for pulsar and neutron star birth spins would involve the spin-up of the initial  spin of the progenitor Chandrasekhar core.  Here, an initial spin of $\sim$50--100 seconds would naturally map into a final spin after collapse and PNS cooling of $\sim$50--100 milliseconds \citep{Ott2006}, a factor of  order a thousand.  Such initial spin periods naturally arise from the application of a Taylor-Spruit dynamo during massive-star evolution \citep{Heger2005} and $\sim$50--100 milliseconds is a reasonable pulsar birth period \citep{Faucher2006}. However, evidence from the asteroseismology of red giants has revealed internal stellar rotation rates that are as much as a factor of ten slower than predicted by such a dynamo \citep{Cantiello}. This suggests that a more efficient mechanism for angular momentum transport out of stellar cores may be in operation, perhaps generically.
This possibility revives the suggestion by \citet{Blondin2007a} that the birth spin is {\it induced} during the post-bounce phase by a spiral SASI (standing accretion shock instability).
This idea was explored in more detail and with a more sophisticated 3D hydrodynamic treatment of the entire PNS core, but with simplistic neutrino physics, by \citet{Rantsiou2011}.
Those authors did not see the spiral SASI, but concluded that spins could indeed be induced. However, the induced spins were no faster than a few seconds.
More recently, using a 3D approach more sophisticated than that of  \citet{Rantsiou2011}, but less sophisticated than that which can now be achieved, \citet{Wongwathanarat2013} derived induced spins periods of 0.1--8.0 seconds, and also explored the associated kick speeds.
In addition, \citet{Nakamura2021} found that a 3D SN1987A model experienced an induced spin of $\sim$0.1 seconds and a kick of $\sim$70 km s$^{-1}$.
Both these papers looked for spin-kick correlations, finding none. Finally, when simulating low-mass electron-capture supernova progenitors in the range $\sim$8.8--9.6 M$_{\odot}$ in 3D, using even better tools, \citet{Stockinger2020} found induced spins of $\sim$30 ms, along with kicks of $\sim$25--40 km s$^{-1}$.
However, though these authors conducted simulations to late times, $\sim$0.5 seconds after bounce they replaced their full transport module (albeit using the dimensionally-reduced  ray-by-ray+ approach) by a much simpler spherical emission scheme that turned off acceleration by anisotropic neutrino emissions. They thereby artificially halted any further neutrino radiation accelerations of the PNS core.

Therefore, the time is ripe for a return to this general issue of the potential for induced spin to contribute to the birth spin of the residue of collapse. Does the \citet{Cantiello} measurement suggest that angular momentum can be effectively drained from the core of a massive star, thereby leaving it at collapse rotating very slowly? And can the subsequently induced rotation by asymmetric matter accretion onto the PNS core after explosion, that also entails stochastic accretion of angular momentum, explain the observed birth periods. We admit that we find this difficult to believe for the generic case and still suspect the initial rotation of the Chandrasekhar core is a factor in the final spin outcome.  However, as we will show in this paper the induced spins for initially non-rotating progenitor cores can be interesting and in the observed range of many inferred pulsar birth spins, again, even for initially non-rotating models.

This paper on PNS kicks and induced spins is a continuation of the long series we have published on the results and insights garnered from a subset of our recent state-of-the-art 3D radiation/hydrodynamic simulations using our CCSN code \fornax \citep{Skinner2019}. In \citet{Radice2019} and \citet{Burrows2019}, we published 3D explosion models for low-mass progenitors, in \citet{Burrows2020} we explored 3D supernova models for progenitors from 9.0 to 25 M$_{\odot}$, in \citet{Radice2019} we presented the corresponding gravitational-wave signatures, and in \citet{Nagakura2020b} we published the corresponding neutrino signatures. Furthermore, in \citet{Vartanyan2019b} we explored the temporal and angular variations in those neutrino emissions, in \citet{Vartanyan2020} we calculated the gravitational-wave signatures due to these time-changing and asymmetrical neutrino fluxes, in \citet{Nagakura2019} we investigated the resolution-dependence of the outcomes of 3D simulations, in \citet{Nagakura2020} we studied the character of inner PNS convection for many of these progenitor models, and in \citet{Vartanyan2021} we investigated the 3D collapse of 3D initial models \citep{Fields2020,Fields2021,Muller2016b,Muller2017}.

For this kick and spin study, we mine the detailed 3D \fornax models published in \citet{Burrows2020} using the Blue Waters platform.  To these models, we have added recent simulations generated at TACC/Frontera \citep{Stanzione2020} and ALCF/Theta.  In \S\ref{models}, we list the simulation models included in this paper. In \S\ref{sec:kicks}, we explain the methods we use to determine the kick magnitudes and directions, due to both matter ejecta and neutrino recoils. In \S\ref{sec:spin}, we summarize the corresponding calculation of the induced spin and then in \S\ref{sec:derivedkicks9} we present our kick results for the 9.0 M$_{\odot}$ progenitor model performed on Frontera to $\sim$1.7 seconds after bounce.  This turned out to be long enough to asymptote in both spin and kick and is our second longest 3D simulation among $\sim$50 performed to date. In \S\ref{sec:derivedkicks}, we expand the discussion of the kick results to the other models, with an aside in \S\ref{black} on black hole kicks. We next transition in \S\ref{sec:inducedspins} to a discussion of the induced spin results.  Finally, in \S\ref{conclusions} we provide a summary of all the salient insights to emerge from this PNS spin and kick investigation and conclude with some general thoughts.

\section{Methods}
\subsection{Simulation Models}
\label{models}

\begin{table*}
    \centering
    \tabcolsep=0.13cm
    \begin{tabular}{ccccccccccccccc}
$M_{\rm ZAMS}$ & Expl. & Machine & $t_{\rm max}$ & $M_{\rm PNS}$ &
$L_{\rm PNS}/10^{46}$ & $P_{\rm PNS}^{\rm(hot)}$ & $P_{\rm PNS}^{\rm(cold)}$ & $\dfrac{T}{|W|}$ &
$v_{\rm kick}^{\rm(matter)}$ & $v_{\rm kick}^{\rm(\nu)}$ & $v_{\rm kick}^{\rm(total)}$ &
$a_{\rm kick}^{\rm(matter)}$ & $a_{\rm kick}^{\rm(\nu)}$ & $a_{\rm kick}^{\rm(total)}$ \\[5pt]
$[M_{\sun}]$ & & & $[{\rm s}]$ & $[M_{\sun}]$ & $[{\rm g\ cm^2\ s^{-1}}]$ &
$[{\rm ms}]$ & $[{\rm ms}]$ & $[10^{-7}]$ &
$[{\rm km\ s^{-1}}]$ & $[{\rm km\ s^{-1}}]$ & $[{\rm km\ s^{-1}}]$ &
$[{\rm km\ s^{-2}}]$ & $[{\rm km\ s^{-2}}]$ & $[{\rm km\ s^{-2}}]$\\
\hline
\bf{9.0} & \bf{yes} & \bf{F} & \bf{1.77} & \bf{1.34} & \bf{1.04} & \bf{1120} & \bf{407} & \bf{1.26} & \bf{71.6} & \bf{91.6} & \bf{142} & \bf{6.39} & \bf{22.5} & \bf{28.7}\\
9.0 & yes & $\Theta$ & 1.42 & 1.34 & 0.710 & 1780 & 553 & 0.567 & 51.5 & 72.9 & 115 & 8.39 & 48.4 & 56.5\\
9.0 & yes & BW & 1.02 & 1.34 & 1.06 & 1330 & 347 & 1.20 & 30.7 & 70.7 & 65.9 & 5.92 & 43.0 & 44.8\\
10.0 & yes & $\Theta$ & 0.605 & 1.48 & 1.33 & 1580 & 255 & 1.16 & 5.88 & 75.5 & 73.4 & 151 & 407 & 413\\
10.0 & yes & BW & 0.767 & 1.49 & 1.48 & 1240 & 260 & 1.53 & 3.46 & 95.2 & 92.9 & 128 & 465 & 357\\
11.0 & yes & $\Theta$ & 0.493 & 1.45 & 2.52 & 925 & 122 & 4.11 & 25.0 & 108 & 86.9 & 237 & 433 & 312\\
12.0 & no & $\Theta$ & 1.17 & 1.69 & 3.46 & 509 & 136 & 6.29 & 0.283 & 65.9 & 65.6 & 2.38 & 563 & 561\\
12.0 & yes & BW & 0.903 & 1.51 & 12.8 & 137 & 31.2 & 113 & 56.2 & 348 & 293 & 251 & 482 & 389\\
13.0 & no & BW & 0.771 & 1.76 & 3.53 & 654 & 125 & 5.07 & 0.130 & 33.5 & 33.4 & 53.6 & 773 & 720\\
14.0 & no & BW & 0.993 & 1.82 & 4.42 & 469 & 111 & 7.86 & 0.165 & 50.1 & 49.9 & 4.28 & 522 & 520\\
15.0 & no & $\Theta$ & 1.09 & 1.94 & 6.84 & 310 & 75.7 & 15.8 & 0.298 & 75.1 & 74.8 & 9.84 & 374 & 370\\
15.0 & no & BW & 0.993 & 1.77 & 4.28 & 459 & 109 & 8.12 & 0.252 & 46.8 & 46.6 & 3.65 & 679 & 678\\
16.0 & yes & BW & 0.616 & 1.58 & 2.79 & 823 & 134 & 4.15 & 45.0 & 140 & 117 & 406 & 399 & 371\\
17.0 & yes & BW & 0.649 & 1.61 & 3.55 & 650 & 110 & 6.40 & 61.1 & 166 & 109 & 535 & 656 & 674\\
18.0 & yes & BW & 0.618 & 1.60 & 3.00 & 778 & 126 & 4.59 & 30.5 & 225 & 197 & 311 & 690 & 401\\
19.0 & yes & BW & 0.871 & 1.75 & 7.46 & 291 & 60.6 & 23.8 & 63.3 & 416 & 353 & 153 & 735 & 640\\
20.0 & yes & $\Theta$ & 0.646 & 1.98 & 4.62 & 676 & 94.4 & 5.49 & 65.6 & 103 & 50.6 & 704 & 687 & 154\\
20.0 & yes & BW & 0.629 & 1.88 & 4.22 & 689 & 104 & 5.43 & 21.4 & 156 & 140 & 276 & 931 & 976\\
25.0 & yes & BW & 0.616 & 1.99 & 6.05 & 521 & 75.7 & 9.31 & 13.6 & 82.2 & 82.9 & 934 & 739 & 993
    \end{tabular}
    \caption{
    Results for the simulations presented in this work.
    $M_{\rm ZAMS}$ is the initial ZAMS mass of the progenitor model.
    The ``Expl.'' column denotes whether the model successfully explodes. The Machine column denotes which HPC cluster the model was run on (BW=Blue Waters, F=Frontera, $\Theta$=Theta); all of the Blue Waters runs were presented in \citet{Burrows2020}.
    $t_{\rm max}$ is the maximum post-bounce time reached by the model.
    $M_{\rm PNS}$ is the residual baryonic PNS mass at $t_{\rm max}$.
    $L_{\rm pns}$ if the angular momentum of the PNS at the end of the calculation.
    $P_{\rm PNS}^{\rm(hot)}$ is the PNS spin period of the end of the calculation (while the PNS is still hot), and $P_{\rm PNS}^{\rm(cold)}$ is the inferred final spin period of the cold final NS state, assuming Eqn.~\ref{eqn:p_cold}. $T/|W|$ is the ratio of rotational kinetic energy to gravitational potential energy at the end of each simulation; note that we have normalized this column by $10^{-7}$ (i.e. gravitational energy dominates by a factor of $\sim10^7$).
    $v_{\rm kick}^{\rm(matter)}$, $v_{\rm kick}^{\rm(\nu)}$, and $v_{\rm kick}^{\rm(total)}$ are the final kick speeds induced by matter/hydrodynamics only, neutrinos only, and the combination of those two, respectively. Similarly, $a_{\rm kick}^{\rm(matter)}$, $a_{\rm kick}^{\rm(\nu)}$, and $a_{\rm kick}^{\rm(total)}$, are the final accelerations for those three cases.
    }
    \label{tab:sims}
\end{table*}

To investigate the magnitude, systematics, and character of the kicks and spins imparted to
the PNS residue of collapse, we data-mine a subset of the large set of 3D models we have recently generated
using our multi-dimensional, multi-group radiation
hydrodynamics code \fornax \citep{Skinner2019}.
The original progenitor models were taken from \citet{Sukhbold2016}
and \citet{Sukhbold2018}. The corresponding massive-star progenitor models have ZAMS masses of 9.0 (F, $\Theta$, BW), 10.0 ($\Theta$, BW), 11.0, 12.0 ($\Theta$, BW), 13.0, 14.0, 15.0 ($\Theta$, BW), 16.0, 17.0, 18.0, 19.0, 20.0 ($\Theta$, BW), and 25.0 M$_{\odot}$, where the letters in parentheses refer to the machine on which the corresponding calculations were performed. If there is no parenthetical machine, Blue Waters (BW) should be assumed; the original paper for all the Blue Waters runs is \citet{Burrows2020}.
Note that some of the Blue Waters models were resimulated on ALCF/Theta ($\Theta$) and that the longest-duration 9.0 M$_{\odot}$ model was done on Frontera (F). The resolution of the Blue Waters runs was 678($r$)$\times$128($\theta$)$\times$256($\phi$), while the Frontera and Theta runs were simulated with resolutions of 1024($r$)$\times$128($\theta$)$\times$256($\phi$).
All of these models were run with 12 logarithmically-distributed energy groups for each of our three neutrino species ($\nu_e$, $\bar{\nu}_e$, and the rest bundled as ``$\nu_\mu$'').
We suggest that providing results for the same progenitor simulated on different machines and at slightly different resolutions can provide the reader with a flavor for some of the possible model variations.
Note also that because the low-mass 9.0 M$_{\odot}$ (F) model asymptoted in explosion energy, PNS mass, kick, and spin, we focus more on this model in the discussions.
As with most simulations in the literature, many of the models we present here did not asymptote.
Nevertheless, the results we find shed light on general behavior and correlations, as well as on the mechanisms that lead to kicks and induced spins.
We note that some of our models do not explode (13(BW), 14(BW), 15(BW), 12($\Theta$), and 15($\Theta$)).
These can illuminate the kicks expected for stellar-mass black holes (\S\ref{black}, see also \citet{Janka2017}).

\subsection{Kick Calculation Method}
\label{sec:kicks}

Allowing the core actually to move from the center of the grid is often computationally fraught. Therefore, many 2D and 3D supernova calculations are performed either with a reflecting boundary condition in the grid center or in 1D (spherical symmetry) interior to a given radius. The former approach is taken in the \fornax calculations presented here, while the latter constraint is otherwise almost universal \citep{Scheck2006,Wongwathanarat2013,Stockinger2020}. Exceptions are few, and are generally in the context of Cartesian AMR grids \citep{Couch2013,Nordhaus2010,Nordhaus2012}.  In fact, the only published calculations that actually witnessed bulk core motion and manifest kicks in collapse calculations were those of \citet{Nordhaus2010} and \citet{Nordhaus2012}, using the AMR code CASTRO \citep{2010ApJ...715.1221A}.  To perform such a calculation, these authors expanded the gravitational potential around the center of mass of the inner core  at densities higher than 10$^{12}$ gm cm$^{-3}$.  A more rigorous, though similar scheme, has been implemented by \citet{Couch2013}. This was necessary, since maintaining the point around which the multipole calculation of the gravitational potential is calculated at the grid center requires extreme precision in the higher-order gravitational multipoles, in particular the dipole.  This is true, despite the possibility that the matter distribution could still be spherical, but around a significantly displaced point.

Hence, in order to estimate kick accelerations and speeds, researchers and papers have traditionally integrated the momentum flux into the PNS core, and the net gravitational ``tug'' on it by external matter that may be aspherically distributed, using the equations of momentum conservation.  We do the same, identifying each component in turn, but use a more sophisticated 3D radiative transfer scheme and neutrino-matter momentum coupling.

As stated in \citet{Vartanyan2019a} and \citet{Skinner2019}, our 3D core-collapse supernova code F{\sc{ornax}} uses the M1 closure to truncate the radiation moment hierarchy by specifying the second
and third moments as algebraic functions in terms of the zeroth and first \citep{vaytet:11}. The basic equations of radiative transfer in the comoving frame that we solve are the zeroth- and first-moment equations of the full equation of radiative transfer for the specific intensity. We follow the evolution of $\nu_e$, $\bar{\nu}_e$, and $``\nu_\mu"$ neutrinos, where the latter represents
the $\nu_{\mu}$, $\bar{\nu}_{\mu}$, $\bar{\nu}_{\tau}$, and $\nu_{\tau}$, neutrinos collectively. Dropping for
clarity the corrections for general relativity\footnote{In fact, the use of the Newtonian momentum equation introduces only small errors
and has been standard past practice \citep{Rantsiou2011,Wongwathanarat2013,Stockinger2020}, 
although see Eqns. 8.6 and 8.13 of \citet{shibata2011} for a concise general-relativistic description.
}, the matter and radiation momentum conservation equations are:

\begin{gather}
\begin{split}
(\rho v_j)_{,t} + (\rho v^i v_j + P \delta^i_j)_{;i} ={} -\rho \phi_{,j}\\
+ c^{-1} \sum_s \int_0^\infty (\kappa_{s\varepsilon} + \sigma^{\rm tr}_{s\varepsilon}) F_{s\varepsilon j} d\varepsilon
\label{new.eq}
\end{split}\\
\begin{split}
F_{s\varepsilon j,t} + (c^2 P_{s\varepsilon j}^i + v^i F_{s\varepsilon j})_{;i} + v^i_{;j} F_{s\varepsilon i} - v^i_{;k} \frac{\partial}{\partial\varepsilon} (\varepsilon Q^k_{s\varepsilon ji}) ={}\\
-c(\kappa_{s\varepsilon} + \sigma^{\rm tr}_{s\varepsilon}) F_{s\varepsilon j},
\end{split}
\end{gather}
where differentiation is indicated with standard notation, $P=P(\rho,e,Y_e)$ is the pressure, $\rho$ is the mass density,
$v_i$ are the velocity components, $\kappa_{s\varepsilon}$ and
$\sigma^{\rm tr}_{s\varepsilon}$ are the absorption and transport scattering opacities, $s\in\{\nu_e,\bar{\nu}_e,``\nu_\mu"\}$, $\varepsilon$ is the neutrino energy, $F_{s\varepsilon j}$ is radiation flux
spectrum (first moment),  $P_{s\varepsilon i}^j$ is the radiation pressure tensor (second moment), $Q^k_{s\varepsilon ji}$
is the heat tensor (third moment), and the other variables have their standard meanings. $\phi$ is the gravitational potential. A comprehensive set of neutrino-matter interactions is
implemented into F{\sc{ornax}}, and these are described in
\cite{Burrows2006}. As formulated, these equations explicitly conserve total momentum (when transforming the neutrino fluxes as we do for this study to the laboratory frame). To calculate the total recoil of the PNS core due to anisotropic neutrino losses we have merely to integrate the flux vectors exterior to the residual PNS core over all solid angles. This is what we do.

Integrating each of the terms in the matter momentum eq. (\ref{new.eq})  over a sphere containing the PNS yields the accelerations on the core and time-integrating these terms provides the kicks induced by these different components. The instantaneous vector sum of these terms gives the total resultant kick, and its asymptotic value yields the supernova-induced final kick.
The individual terms are:
\begin{align}
\label{eqn:accel_mom}
    \dot{\mathbf{p}}_{\rm mom}(R)&\equiv -R^2\oint_{r=R}\rho\mathbf{v}\otimes\mathbf{v}\cdot\mathbf{n}\id\Omega\\
\label{eqn:accel_pres}
    \dot{\mathbf{p}}_{\rm pres}(R)&\equiv -R^2\oint_{r=R}P\,\mathbf{n}\id\Omega\\
\label{eqn:accel_grav}
    \dot{\mathbf{p}}_{\rm grav}(R)&\equiv \int_{r\ge R}\rho\nabla_{\rm cart}\phi\id{V}\\
\label{eqn:accel_nu}
    \dot{\mathbf{p}}_{\nu_i}(R)&\equiv -R^2\oint_{r=R}
    \dfrac{\mathbf{F}_{\nu_i}\otimes\mathbf{n}}{c}\cdot\mathbf{n}\id\Omega,\\
    \dot{\mathbf{p}}_{\rm matter}(R)&\equiv\dot{\mathbf{p}}_{\rm mom}(R)+\dot{\mathbf{p}}_{\rm pres}(R)+\dot{\mathbf{p}}_{\rm grav}(R)\\
    \dot{\mathbf{p}}_{\nu}(R)&\equiv\sum_i \dot{\mathbf{p}}_{\nu_i}(R)\, ,
\end{align}
where $\mathbf{n}=\left\{\sin\theta\cos\phi,\sin\theta\sin\phi,\cos\theta\right\}$ is the unit normal to the sphere,
$\nabla_{\rm cart}=\left\{\partial_x,\partial_y,\partial_z\right\}$ is the Cartesian gradient, and $\phi$ is the gravitational potential.
To clarify, the integrand in Eqn.~\ref{eqn:accel_nu} is $\mathbf{F}^i\mathbf{n}^j\mathbf{n}^i$ when written in standard Einstein notation (omitting the subscript $\nu_i$).
The ``matter" component is the sum of the first three of these integrated forces (Eqn.~\ref{eqn:accel_mom}-\ref{eqn:accel_grav}), and the neutrino component is the sum over all the neutrino species $i$ of Eqn.~\ref{eqn:accel_nu}.
The term $\dot{\mathbf{p}}_{\rm g}(R)$ is the so-called gravitational ``tugboat'' effect. Since many kick estimates are halted long before the final kicks are achieved and since early during the asymmetrical ejection of the supernova debris the gravitational effect can be large, this term has at times been given a disproportionate emphasis, to the degree that the theoretical supernova kick mechanism is oftimes referred to as the gravitational ``tugboat'' mechanism. However, this is misleading, since the resulting core motion is in fact due merely to global momentum conservation; the kicks in the end result solely from final-state matter and neutrino recoil.

For some of the discussion below, we find it convenient to define instantaneous and running average anisotropy parameters:
\begin{align}
\label{eqn:aniso1}
    \alpha_j&=\left.\left|\dot{\mathbf{p}}_{j}\right|\middle/\left[\oint_{r=R}
    \left|\pder{\dot{\mathbf{p}}_{j}}{\Omega}\right|\id\Omega\right]\right.\\
\label{eqn:aniso2}
    \left<\alpha_j\right>_t\left(t\right)&=\left.\left|\int_0^t\dot{\mathbf{p}}_{j}\id\tau\right|\middle/\left[\int_0^t \oint_{r=R}
    \left|\pder{\dot{\mathbf{p}_{j}}}{\Omega}\right|\id\Omega\id\tau\right]\right.
\end{align}

\begin{align}
    \alpha_{\rm mom}(R,t)&\equiv \left.\left|\dot{\mathbf{p}}_{\rm mom}\right|\middle/\left[R^2\oint_{r=R}\rho\mathbf{v}\cdot\mathbf{v}\id\Omega\right]\right.\\
    \alpha_{\rm pres}(R,t)&\equiv \left.\left|\dot{\mathbf{p}}_{\rm pres}\right|\middle/\left[R^2\oint_{r=R}P\id\Omega\right]\right.\\
    \alpha_{\rm grav}(R,t)&\equiv \left.\left|\dot{\mathbf{p}}_{\rm grav}\right|\middle/\left[\oint\left|\int_{r\ge R}\rho\nabla_{\rm cart}\phi R^2\id{r}\right|\id{\Omega}\right]\right.\\
    \alpha_{\nu_i}(R,t)&\equiv \left.\left|\dot{\mathbf{p}}_{\nu_i}\right|\middle/\left[R^2\oint_{r=R}
    \dfrac{\left|\mathbf{F}_{\nu_i}\cdot\mathbf{n}\right|}{c}\id\Omega\right]\right. \, .
\label{aniso}
\end{align}
$R$ is the radius at which we calculate the momentum fluxes to the core.  We generally let this radius track the mass coordinate corresponding to $R=100$ km at the final timestep of the simulation, 
while using the isodensity surface of $\rho=10^{11}$~g~cm$^{-3}$ is a common alternative \citep[see e.g.][]{Wongwathanarat2013}
\footnote{
For the 9.0 M$_{\odot}$ (F) model, the resulting kicks differ from those derived using the $\rho=10^{11}$ g cm$^{-3}$ surface (as is frequently done) by of order $\sim10\%$, with no notable qualitative differences. 
We prefer our choice of coordinate since it is well defined before bounce, is pseudo-Lagrangian, and for asymptoted models it guarantees that we capture all the accreted material.
}.
The last anisotropy term in particular, referring to the dipole asymmetry of the neutrino emissions (it would be zero if the latter were isotropic), is the term in Eqn. (\ref{nkick}) that plays such an important role in the neutrino contribution to the kick speed.

\subsection{Spin Calculations}
\label{sec:spin}

After explosion, the asymmetrical accretion of matter in downward plumes
onto the core can continue for some time.  These plumes not only excite core oscillations that lead to gravitational-wave radiation of significance \citep{Radice2019}, but bring in angular momentum, whose sign depends upon
a stream's given impact parameter. The accretion of angular momentum is thereby quasi-stochastic and there is initially an apparent degree of randomness in the magnitude and sign of each accreted blob. The ``wasp-waist'' \citep{Burrows2021} pinched character of the early accreta, so important for continuing the neutrino driving after the onset of explosion, is one aspect of this phenomenon.  However, the fallback is also affected by the asymmetry in the explosion itself, and determining this requires the detailed calculations we have conducted.  Moreover, it is the outer shells of the PNS that are spun up $-$ the transfer of the accreted angular momentum to the inner core is affected on long viscous and/or magnetic torquing timescales.  Our cores do not, therefore, achieve solid-body rotation during the timescales of our simulations (see Table \ref{tab:sims}).  Therefore, the spin rates and periods we calculate, tabulate, and plot in Table \ref{tab:sims} and \S\ref{sec:inducedspins} are those distributing the total accreted angular momentum to create a solid-body rotator with the same density distribution.
We also provide the final spin rate and period calculated by taking the accreted angular momentum at the end of each simulation and artificially contracting the core mass interior to a density isosurface of 10$^{11}$ g cm$^{-3}$ to an outer radius of 12 km\footnote{While this implies that the moment of inertia is proportional to $R_{\rm PNS}^2$, the uncertainty in this assumption is likely less than that in assuming $R_{\rm PNS}=12$ km.}. The formula employed is:
\begin{align}
\label{eqn:p_cold}
    P_{\rm PNS}^{\rm(cold)}=\left(\dfrac{R_{\rm PNS}}{12\ {\rm km}}\right)^{2}P_{\rm PNS}^{\rm(hot)}\, .
\end{align}
Twelve kilometers is a reasonable canonical neutron star radius after prolonged deleptonization and cooling (which has only just begun by the end of our simulations). The PNS radii at the end of our simulations are generally 25$-$30 km, so there is anticipated to be significant further spin-up due to continuing contraction.  However, by the end of many of our exploding calculations the accretion phase is starting to taper off, but has not terminated. For the 9.0-M$_{\odot}$ model it has effectively stopped.

Finally, we comment that the chaotic and turbulent nature of 3D supernova explosions
implies that a given progenitor will realistically give rise to a spectrum of outcomes. The explosion energies, residue baryon masses, nucleosynthesis, blast morphologies, spins, and kicks will be drawn from distribution functions that are not delta functions, though some of them could be narrow.  However, finite ``widths'' are  universal and physically realistic expectations \citep{Burrows2021}, though these distributions are currently unknown. Determining them will entail much future work and is an important community goal going forward.

\section{Results}
\subsection{Kicks}

\subsubsection{Kick for the Long-term 9.0 \texorpdfstring{M$_{\odot}$}{Mo} Progenitor}
\label{sec:derivedkicks9}

The 3D 9.0 M$_{\odot}$ model calculated on Frontera (F) was carried to 1.774 seconds after bounce, a time sufficient to witness the asymptoting of many of the properties of the explosion and the birth of the neutron star residue.  These include the PNS baryon mass (1.34 M$_{\odot}$), the explosion energy ($0.107\,\times\,10^{51}$ ergs $=0.107$ Bethe), the kick, and the induced spin (\S\ref{sec:inducedspins}). For full-physics 3D models in the literature, this is rare, since most exploding models require many seconds to achieve this state \citep{Muller2017,Burrows2021} and are, hence, too expensive to obtain using most current computational tools and platforms.  The steeper massive-star progenitor density
profile of this 9.0 M$_{\odot}$ star \citep{Sukhbold2016} results in an earlier explosion and speeds up the subsequent evolution. In the context of the explosion-induced kick,  we see this in Fig. \ref{fig:9.0_kick_evo}, which depicts the development with time of the magnitudes of the various components of the acceleration of the PNS core (left) and the magnitudes of the corresponding components to the resulting kick (right) (its integral quantity).  The black curve is the magnitude of the vector sum of all the contributions, the orange line is the summed effect of the matter terms (see \S\ref{sec:kicks}), and the blue line is the magnitude of the vector sum of the neutrino contributions. We see that after $\sim$1.6 seconds the accelerations have subsided (left) and the kicks have saturated (right).  Both panels of Fig. \ref{fig:9.0_kick_evo} provide the z-components (dotted) of the various terms. The latter are merely to provide on this plot a sense of the relative direction $-$ though the x- and y-components are calculated they are not given here. Of course, the total magnitude includes their vector contribution. 
In Fig.~\ref{fig:kick_methods}, we show how the resulting kicks would be affected if we instead we chose the $\rho=10^{11}$ g cm$^{-3}$ isosurface. The end state kicks are modified at the $\sim$10\% level, with minimal qualitative differences.

The fact that the terms on the left side of Fig. \ref{fig:9.0_kick_evo} taper to near zero is the singular aspect of this calculation.  Previous calculations merely grew and fluctuated, but did not subside to reveal the final kick, nor the role of each contribution over a full development cycle. What we find is that for this model the total matter and neutrino kicks roughly add, with the neutrino effect being the larger.  The asymptotic total kick magnitude is $\sim$142 km s$^{-1}$ (see Table \ref{tab:sims}), below the mean kick speeds generally measured.  This lower value is expected for an initially spherical progenitor  that explodes early and with low explosion energy \citep{Burrows2007,Janka2017}.
Such a model doesn't have much time to develop very vigorous neutrino-driven convection after bounce before it explodes, with the result that the various terms that give rise to the kick (\S\ref{sec:kicks}) don't evolve large dipoles. This is particularly true for the matter terms, which for this model don't sum to greater than an acceleration of $\sim$100 km s$^{-2}$.

Interestingly, as Fig. \ref{fig:9.0_detailed_kick_evo} demonstrates, the gravitational term (blue ``tugboat'') and the pressure term (green) nearly cancel, while the ``momentum'' term (orange) remains small.  Moreover, in this model the $\nu_e$ and $\bar{\nu}_e$ neutrino kicks roughly cancel, leaving the ``$\nu_{\mu}$'' component, made up of the $\nu_{\mu}$, $\bar{\nu}_{\mu}$, $\nu_{\tau}$, and $\bar{\nu}_{\tau}$ neutrinos collectively, to assume the primary role.  Figure \ref{fig:9.0_anisotropy} depicts the instantaneous dimensionless neutrino anisotropy parameters, as defined in \S\ref{sec:kicks}.  They grow over a period of $\sim$400 milliseconds, then flatten, and at $\sim$1.2$-$1.4 seconds start to decline. The peak magnitudes for the $\nu_e$ and $\bar{\nu}_e$ neutrinos are $\sim$8$-$10\%, but for this 9.0 M$_{\odot}$ model they are of opposite sign. The net effect is closer to that for the ``$\nu_{\mu}$'' neutrinos at $\sim$1\%.

Rather curiously, the effective duration ($\sim$1 second) for the 9.0 M$_{\odot}$ (F) model of significant neutrino anisotropy is shorter than the expected timescale for the loss of the total binding energy of the cold final neutron star ($\ge$10 seconds); only a fraction of the total neutrino energy radiated (less than 30\% of the total) is radiated here while $\alpha$ is notable.  This may be an important generic aspect of neutrino kicks $-$ the anisotropies are large only for a fraction of the total neutrino emission phase. As a result, one can't merely multiply a respectable anisotropy factor by the total radiated binding energy (divided by $c$) to obtain the neutrino kick. Whether what we see for this long-duration 9.0 M$_{\odot}$ (F) model is generic is to be determined, but the possibility it suggests is intriguing.

For this and other models (\S\ref{sec:derivedkicks}), we generally see a directional correlation between the kicks due to the $\bar{\nu}_e$ and ``$\nu_{\mu}$'' neutrinos. This is not unexpected, since their respective neutrinospheres are deeper than that for the $\nu_{e}$ neutrinos; they are more similarly prone to being distorted by outer accretion plumes and asymmetries. However, though a correlation is seen for this 9.0 M$_{\odot}$ (F) model between the directions of $\nu_{e}$ and $\bar{\nu}_e$ neutrino kick directions, it seems not that generic. We will delve into this question in the next section.

Importantly, unlike the 9.0 M$_{\odot}$ (F) model, for many other models we frequently observe an {\it anti}-correlation between the summed matter kick and net neutrino kick. As with the 9.0 M$_{\odot}$ (F) model, the net neutrino kick frequently assumes the greater role.  This is not in keeping with the spirit of the traditional recoil model, which highlights the net matter effect.  Our tentative conclusion suggests that, at the very least, a hybrid neutrino/matter recoil model might be generic. Since we see evidence of an anti-correlation between the matter and neutrino effects (\S\ref{sec:derivedkicks}), there would still be a relation between the explosion dipole and the kick direction. However, due to the potential importance of the neutrino kick and its frequently oppositely-directed impulse, the net kick direction might be in the opposite direction to what might have earlier been assumed.  This possibility clearly deserves further scrutiny, but is one possibility to emerge from this study.

We note that we had earlier carried out on Theta another 9.0 M$_{\odot}$ simulation, as well as one on Blue Waters (see Table \ref{tab:sims}). The initial model, resolution, and code for the Theta run were the same as on Frontera. The run on Blue Waters used 678 radial zones and a slightly different opacity module. Importantly, in the context of chaotic flow, HPC simulations are not perfectly deterministic.  Slight changes, even at the level of machine precision, or due to very slightly different machine- and compiler-dependent optimization, can diverge in detail.  However, the 9.0 ($\Theta$) and 9.0 (BW) simulations also manifested small kicks ($\sim$115 and $\sim$ 66 km s$^{-1}$, respectively) for which the neutrino component again was comparable to or exceeded in magnitude the net matter component.  As for the Frontera run, for the 9.0 M$_{\odot}$ run on Theta,  the $\nu_e$ and $\bar{\nu}_e$ neutrino kicks were also roughly opposite, leaving the $\nu_{\mu}$ neutrino kick again to dominate.  The matter accelerations settled to very low values within $\sim$1.0 seconds after bounce.  However, that simulation was carried to only $\sim$1.43 seconds (the BW run was carried to only $\sim$1.06 seconds), at which point the neutrino contribution to the kick was just starting to subside (as with the Frontera run).  Nevertheless, the kicks for the three 9.0 M$_{\odot}$ runs are broadly consistent and may hint at the magnitude of variations in the asymptotic states around this progenitor mass.  Furthermore, we note that the final baryon masses for all three runs were nearly identical, as were the explosion time and energy ($\sim$0.106 and $\sim$0.1 Bethes for the Theta and BW runs, respectively).

\subsubsection{Kick Trends for the Other Models}
\label{sec:derivedkicks}

In Table \ref{tab:sims}, we summarize the basic results of this study for the 3D CCSN models included. The numbers provided for each quantity are their final values (at each model's respective $t_{\rm max}$).  Here, we focus on the kicks. Note that only the 9.0 M$_{\odot}$ models have asymptoted in energy and PNS mass and that none of the others have.  For all the other models, what we witness and can report are the results of only the early stages of their evolution, many of which will not achieve their final kick speeds before $\sim$3$-$7 seconds after bounce.  Hence, we view these results as merely suggestive of general correlations and trends.  Also, we emphasize that the kick and acceleration numbers provided are the relevant vector sums.  Since the neutrino and matter components are not aligned, the ``total'' magnitudes given in the Table are the magnitudes of their vector sums.

\begin{figure*}
    \includegraphics[width=\linewidth]{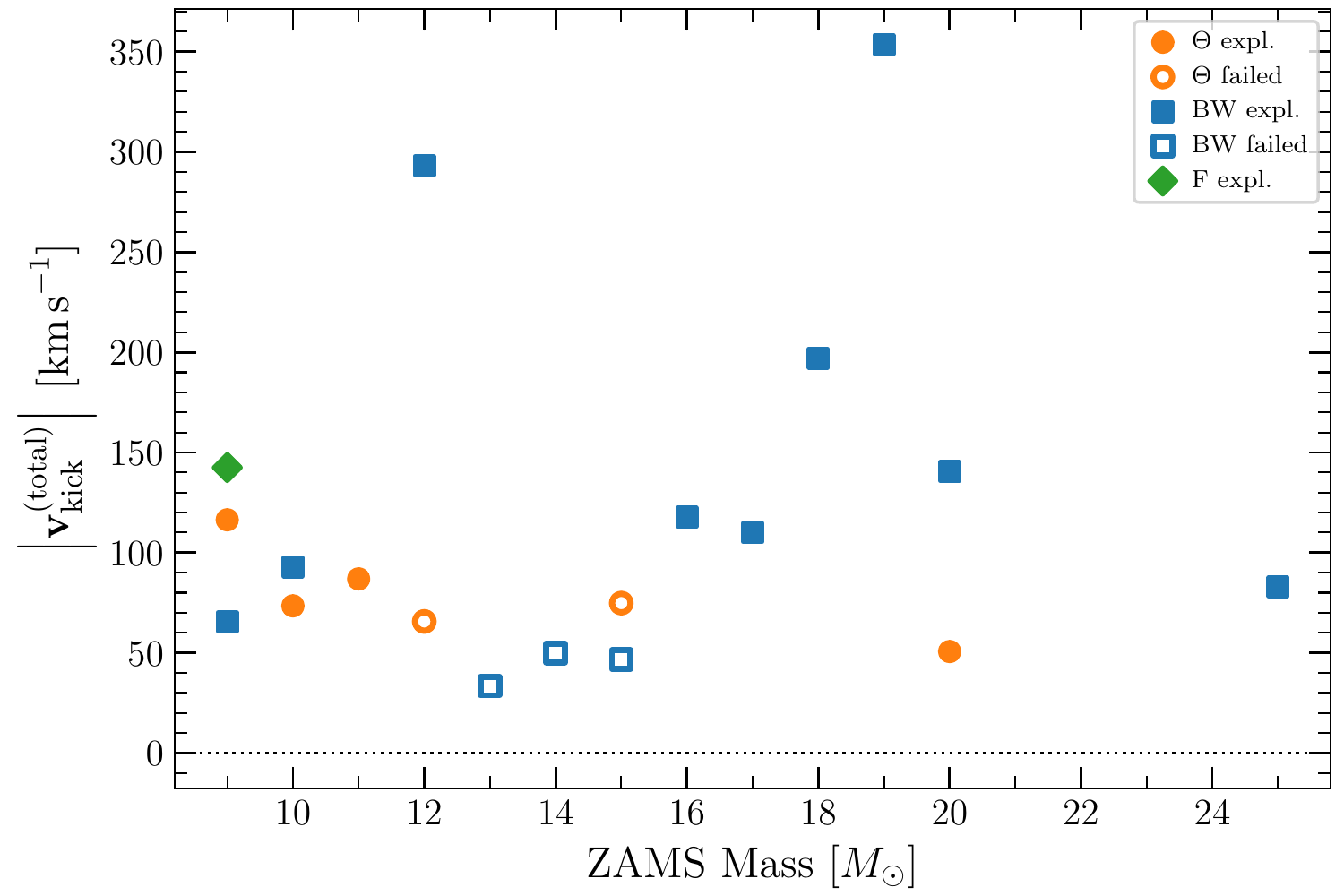}
    \caption{Final PNS kick speed
    as a function of ZAMS mass for our various models.
    The models are colored by which HPC cluster the model was run on (see legend; BW=Blue Waters, F=Frontera, $\Theta$=Theta).
    Open symbols represent models that fail to explode, while filled symbols signify successful explosions.
    We note that this figure caries an implicit assumption that the final mass of the stellar compact remnant is close to that of the residual PNS Baryon mass, which is likely untrue for the non-exploding models. For reference, models with progenitor ZAMS masses of 9.0, 10.0, 11.0,
12.0, 13.0, 14.0, 15.0, 16.0, 17.0, 18.0, 19.0, 20.0, and 25.0 are shown.  See Table \ref{tab:sims} for further information.}
    \label{fig:mass_kick}
\end{figure*}

\begin{figure*}
    \includegraphics[width=\linewidth]{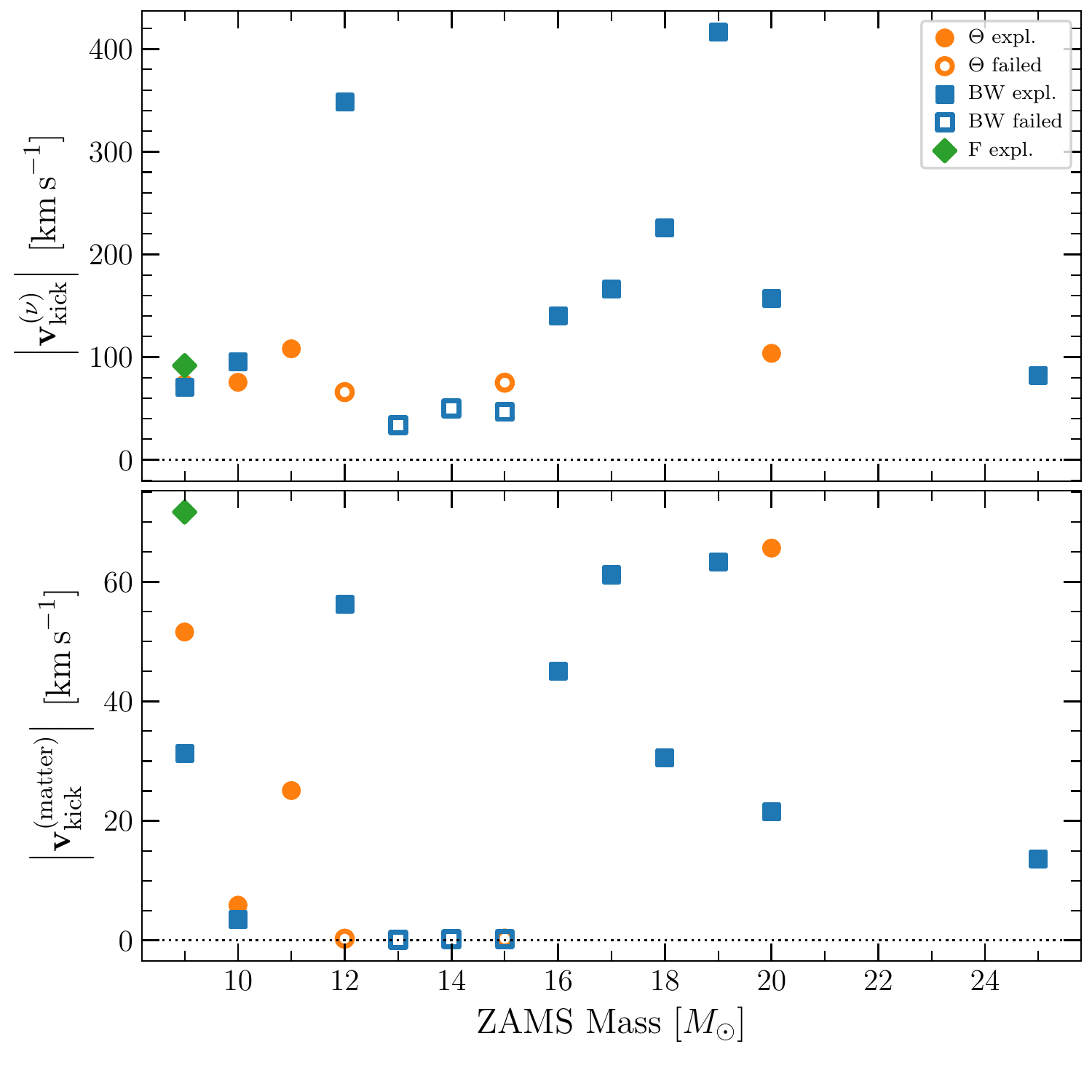}
    \caption{Same as Fig.~\ref{fig:mass_kick}, but showing only the final neutrino component of the kick velocity (top), as well as only the final matter component of the kick velocity (bottom).}
    \label{fig:mass_nu-matter-kick}
\end{figure*}

First, we note that for the non-exploding models the matter kick accelerations and integrated kicks are small, and significantly smaller than the corresponding neutrino kicks.  This is reasonable in the context of explosion duds and is our basic conclusion for the kicks given to black holes birthed in core-collapse. We will discuss this further in \S\ref{black}. Next, we find that for exploding models the kick accelerations due to neutrinos are comparable in magnitude to, and frequently larger than, those due to matter.  This is a new finding to emerge from this study and using 3D simulation results. However, our conclusion needs substantiation and should be viewed as provisional. If true, however, it suggests that the recoil model of pulsar kicks might need revision to incorporate an important role for neutrinos.  Figure \ref{fig:mass_kick} summarizes the total kicks found at the end of each simulation. Included are values for some models done on different machines and using slightly different variants of the opacity tables\footnote{The opacity tables used for the Frontera and Theta models were the same.}. Such variations may hint at the degree of intrinsic variation due to turbulent chaos.
Figure \ref{fig:mass_nu-matter-kick} portrays the dependence found for both the neutrino (top) and matter (bottom) contributions to the kicks as a function of ZAMS mass.  Note the difference in scale.

Figure \ref{fig:vnu_dot_vmat} depicts one of our most intriguing potential findings.  It suggests that for most of our models the matter and neutrino kicks are roughly anti-aligned; they are generally in roughly opposite directions, though not perfectly so.
The major exceptions are the three 9.0 M$_{\odot}$ models, for which the tendency to be anti-aligned is either reversed or muted. This difference may be due to the fact that these models explode earlier than the others, before significant post-shock turbulence can develop. Clearly, more work on this possibility and its robustness is called for, but if true it further emphasizes the importance of the neutrino recoil contribution to pulsar proper motions.

Other interesting observations are encapsulated in Figures \ref{fig:vnu1_dot_vnu2} and \ref{fig:vnu0_dot_vnu1}.  The first figure demonstrates that (at least in these 3D models) the integrated $\bar{\nu}_e$ and $\nu_{\mu}$ neutrino kicks are fairly aligned.
This is not unexpected, since their respective neutrinospheres are both deep.  The second explores the relative alignment of the $\nu_e$ and $\bar{\nu}_e$ neutrino kicks.  While there is some scatter in this dot product, there is a slight indication that the degree of alignment (or anti-alignment) roughly increases with ZAMS mass.
Since there is a very slight tendency for later explosions for those models with the shallowest initial density profiles, and since the more massive progenitors have such profiles, this plot may hint at an interesting rough correlation.  However, this suggestion is not firmly demonstrated by our provisional results and warrants, however intriguing it is, much further study.

The sets of Figures \ref{fig:12.0-oa_kick_evo}, \ref{fig:17.0-oa_kick_evo}, and \ref{fig:19.0-oa_kick_evo}, the sets of Figures \ref{fig:12.0-oa_detailed_kick_evo}, \ref{fig:17.0-oa_detailed_kick_evo}, and \ref{fig:19.0-oa_detailed_kick_evo}, and the sets of Figures \ref{fig:12.0-oa_anisotropy}, \ref{fig:17.0-oa_anisotropy}, and \ref{fig:19.0-oa_anisotropy} provide for the 12-, 17- and 19-M$_{\odot}$ progenitors those time-dependent plots that correspond to Figures \ref{fig:9.0_kick_evo}, \ref{fig:9.0_detailed_kick_evo}, and \ref{fig:9.0_anisotropy} (with minor modifications), respectively. These are provided merely as representative of the models which have not asymptoted, but for which further interesting observations can be made.  We see that the matter accelerations are always delayed with respect to the neutrino accelerations, and, as stated, are generally lower (though not always - see Table \ref{tab:sims} and Figure \ref{fig:mass_nu-matter-kick}).  We also note that the various matter and neutrino terms can partially cancel, so due diligence concerning their vector nature is called for. In addition, it can take $\sim$0.3 to 0.6 seconds after bounce for the neutrino kick asymmetry to peak and roughly flatten.  There is an indication that the longer the time to explosion the longer the time to the flat phase.  This is seen, for example, in the comparison of the 12 M$_{\odot}$ (Figure \ref{fig:12.0-oa_anisotropy}) and 19 M$_{\odot}$ (Figure \ref{fig:19.0-oa_anisotropy}) models. It eventually does flatten, and the behavior of the 9.0 M$_{\odot}$ (F) model in Figure \ref{fig:9.0_anisotropy} may be roughly generic. This suggests that the neutrino asymmetry parameter ramps up quasi-exponentially over a period of many tenths of a second, flattens on average, and then decays, leaving the asymptotic kick speed.  Also, the magnitude of the net neutrino asymmetry parameter can reach $\sim$1$-$10\% for a few seconds, with the different components oftimes partially cancelling.

Finally, we comment that the pressure and gravity components of the matter kicks are typically aligned (though the 9.0 M$_{\odot}$ (F) model has them anti-aligned).  Also, the momentum and gravity components of the kicks are typically anti-aligned (though, interestingly, they are roughly orthogonal for the 9.0 M$_{\odot}$ (F) model).  Furthermore, we see no significant correlations between the neutrino kick direction and the dipole direction of the Lepton-number Emission Self-Sustained Asymmetry \citep[LESA;][]{Tamborra2014,Vartanyan2019b}.

\subsubsection{Theoretical Black Hole Kicks}
\label{black}

Though most of our models explode, our 12 M$_{\odot}$ ($\Theta$), 13 M$_{\odot}$ (BW), 14 M$_{\odot}$ (BW), and 15 M$_{\odot}$ (BW, $\Theta$) models do not.  This outcome is mostly associated with their initial mass density profiles and the evolution of the post-bounce accretion rate through the stalled shock and is the subject of current research \citep{Burrows2021}. For marginal cases, the outcome could also be a function of resolution \citep{Nagakura2019}.  Be that as it may, this collection of models can inform our expectations concerning kicks for black holes.  We see from Table \ref{tab:sims} that for these models the matter kicks are always quite small, while the neutrino kicks are still interesting.  Since for these models an anisotropic explosion of matter was not realized, we would expect the various matter recoil effects outlined in \S\ref{sec:kicks} to be quite small.  This is the case.

Figures \ref{fig:14.0-oa_kick_evo}, \ref{fig:14.0-oa_detailed_kick_evo}, and \ref{fig:14.0-oa_anisotropy} for the non-exploding 14.0 M$_{\odot}$ (BW) model correspond to Figures \ref{fig:9.0_kick_evo}, \ref{fig:9.0_detailed_kick_evo}, and \ref{fig:9.0_anisotropy} and delineate the temporal development of the various contributions to the kick accelerations and speeds for this representative non-exploding model that should evolve over time to a black hole. The basic results are that the matter contributions are quite small, that the neutrino kicks predominate, and that the latter, though still growing, are smaller than for the average exploding model.
However, care should be taken not to over-interpret this result and the model should be continued much further in time.  Nevertheless, the suggestion is that the momentum imparted to the residual core is due to the neutrinos, that the associated anisotropy may be smaller ($\sim$4$-$5\%) than for some of the exploding models, and that the imparted momentum is also smaller than for other models.  Since the model does not explode, the turbulence behind the shock may result in dipole anisotropies of the neutrino emissions that tend to partially average out.  This is certainly true for the net matter component, but also seems true for the net neutrino kick.

The upshot is that our calculations suggest that the imparted linear momentum is not larger ``on average'' than that associated with the corresponding neutrino term for the average exploding model. We, therefore, expect that the net kick for the black hole that results after a $\sim$5$-$15 M$_{\odot}$ helium core or whole star is accreted to yield a $\sim$10 M$_{\odot}$ black hole residue should be $\sim$1.5 M$_{\odot}$/10 M$_{\odot}$ times smaller.  This implies that the resultant $\sim$10 M$_{\odot}$ black hole would recoil due to the net neutrino kick with a speed below $\sim$100 km s$^{-1}$.  Hence, our prediction is that black holes are kicked by neutrinos to smaller speeds than neutron stars birthed in core collapse and that the speeds they reach are determined by the net neutrino recoil, not the net matter recoil.

\subsection{Induced Spins}
\label{sec:inducedspins}

The distribution and evolution of angular momentum in stars is a central topic in stellar astronomy.  Star formation processes impart a star's initial spin. During a star's life, wind mass loss and angular momentum redistribution by various torquing processes accompany its thermal, luminosity, and structural evolution.  The residues of stellar evolution inherit the result.  For a massive star, the initial angular momentum of the interior mass that will eventually collapse when a Chandrasekhar mass is assembled during the star's terminal phase must decrease significantly.  Otherwise, its steady compaction during the various progressive thermonuclear phases would likely result in a severely centrifugally-supported object that could not collapse.  Hence, angular momentum redistribution out of the core is an important aspect of the pre-supernova life of a massive star.

However, the specific processes by which angular momentum is shuttled out of the core and lost via winds at the stellar surface are not known. They likely involve magnetic torques, but the details are as yet unclear.  \citet{Heger2005} have conducted evolutionary calculations of massive stars incorporating a Taylor-Spruit dynamo to transport angular momentum. These resulted in initial core spin periods around $\sim$30$-$60 seconds, which translate via collapse and subsequent neutrino cooling to initial pulsar periods of $\sim$30$-$60 milliseconds \citep{Ott2006}.  To within factors, these spins are reasonable starting spins for radio pulsars \citep{Faucher2006,Kaspi2016}.  However, as \citet{Cantiello} have found in the context of red giants, the use of a Taylor-Spruit angular momentum redistribution scheme can underestimate the true degree of core de-spining by a factor of $\sim$10.   One is left asking the question: Could the efficiency of angular momentum loss from the core be such that the Chandrasekhar core upon collapse is spinning very slowly? And, therefore, could the spin of the pulsar birthed in a supernova explosion be due to stochastic processes around the time of explosion?
This possibility was first broached by \citet{Blondin2007a} and \citet{Blondin2007b}, and then studied by \citet{Rantsiou2011}, \citet{Wongwathanarat2013},  \citet{Stockinger2020}, and \citet{Nakamura2021}.  We ourselves suspect that the initial spin at collapse may be more determinative of initial pulsar spins.  However, whether pulsar spins at birth could be due to the stochastic accretion of matter plumes just before, during, and after explosion remains intriguing and has motivated this study.

\begin{figure*}
    \includegraphics[width=\linewidth]{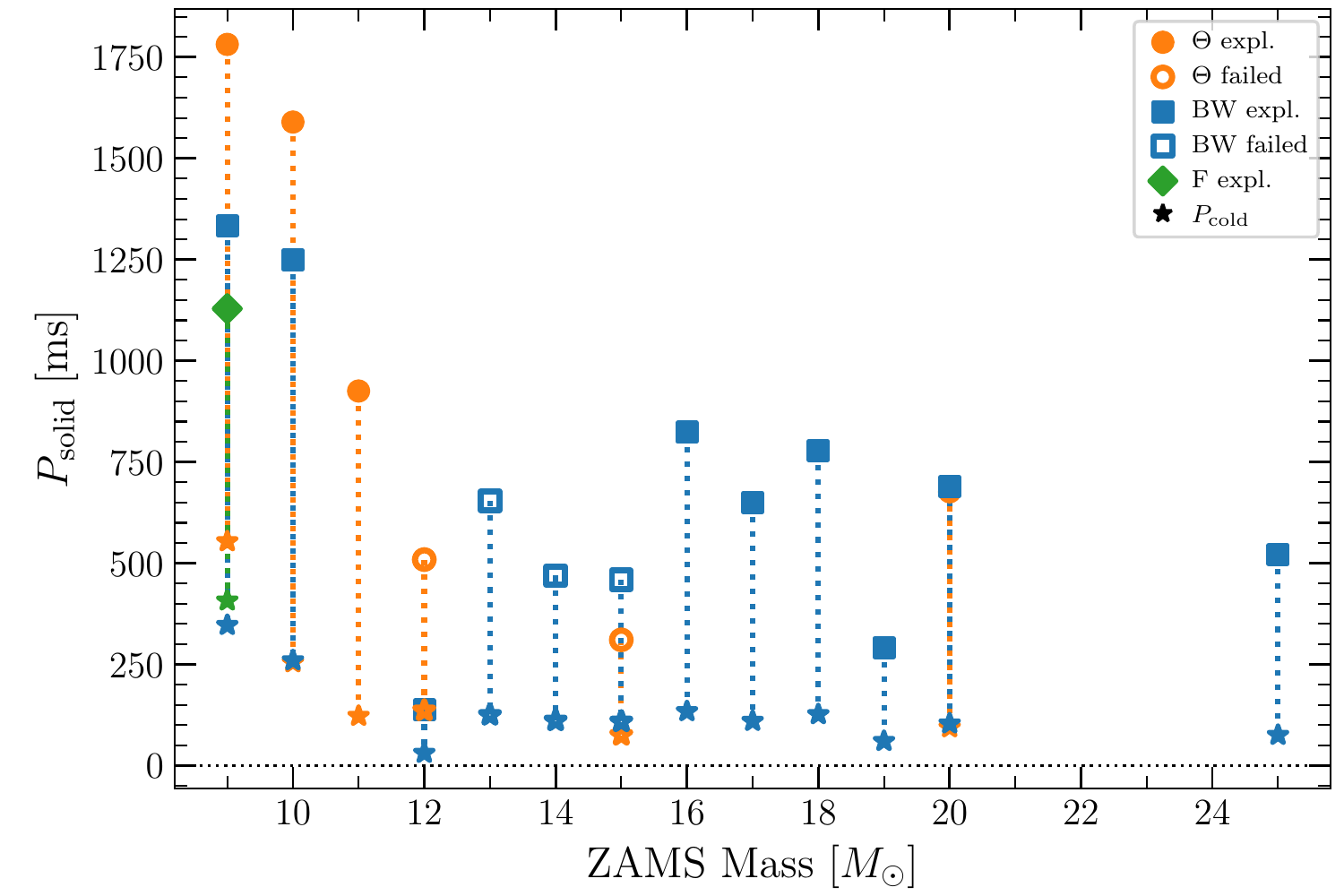}
    \caption{PNS period (assuming solid-body rotation) as a function of ZAMS mass for the various models. The star symbols estimate the final ``cold'' neutron star spin period assuming $P_{\rm cold}=P\times(12\ {\rm km}/R_{\rm PNS})^2$, where $R_{\rm PNS}$ is the mean radius of the $10^{11}$ g cm$^{-3}$ isosurface at the end of the respective calculation. The other colors/symbols have the same meaning as in previous figures (first described in Fig.~\ref{fig:mass_kick}).
    }
    \label{fig:mass_period}
\end{figure*}

Table \ref{tab:sims} provides our results for the angular momenta ($|\vec{L}|$), the spin periods, and the ratio of rotational kinetic energy to gravitational energy ($T/|W|$) at the end of our eighteen detailed 3D simulations, as well as an estimate of the final spin after core deleptonization and cooling.  Figure \ref{fig:mass_period} provides the spin periods (at the end of each simulation and upon achieving the cold, catalyzed state), under the assumption of solid-body rotation at the total angular momentum achieved at the end of each simulation. The vector nature of the total angular momentum was respected, so the final direction of each spin direction was a scatter plot in solid angle.

We see from Figure \ref{fig:mass_period} and Table \ref{tab:sims} that, though the spin period at the end of each simulation hovers around $\sim$500 milliseconds for most models, for the less massive progenitors, in particular the 9.0 M$_{\odot}$ models, the periods are slower than $\sim$1000 milliseconds.  However, when shrunk to outer PNS radii of 12 kilometers, for the final angular momenta achieved in our 3D simulations the spin periods reside in the interval $\sim$30$-$550 milliseconds. Most are near $\sim$100 milliseconds,
and such periods are reasonable birth periods for radio pulsars.  However, we do not view this result as definitive, merely suggestive.
At the very least, these results point to the potential importance of stochastic spin up during the supernova phase.

\begin{figure*}
    \includegraphics[width=1\linewidth]{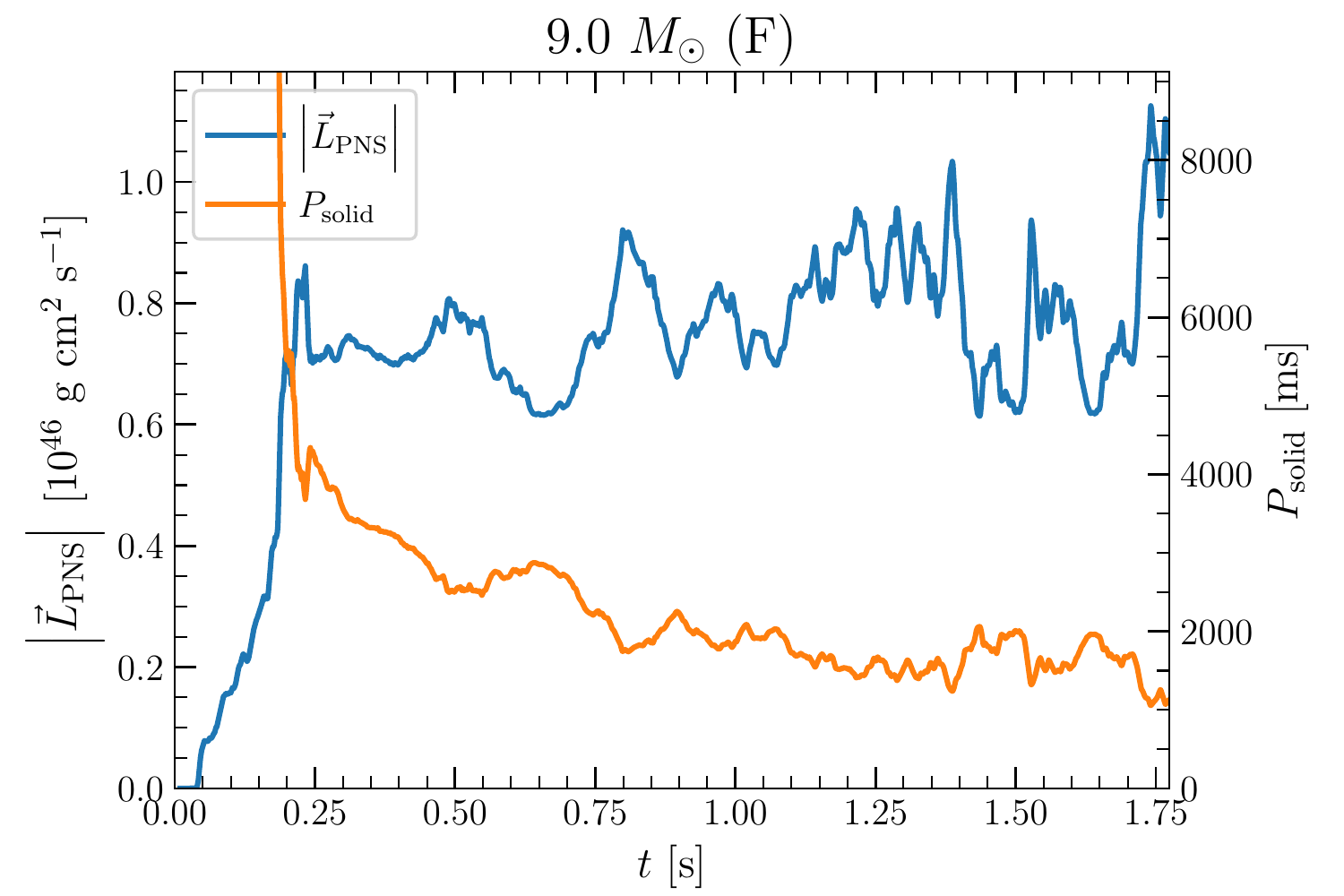}
    \caption{PNS angular momentum (blue; left vertical axis) and spin period assuming solid body rotation (orange; right vertical axis) as a function of time for the 9.0 $M_{\sun}$ (F) model.}
    \label{fig:9.0_spin_evo}
\end{figure*}

Figure \ref{fig:9.0_spin_evo} portrays the evolution versus time after bounce of both the magnitude of the angular momentum and the associated solid-body period for the long-term 9.0 M$_{\odot}$ (F) progenitor simulation. We see that most of the spin-up happens in the early phase before explosion, after which the random accretion of infalling plumes that grow the residual core induces subsequent fluctuations of $\sim$20$-$40\%.  The spin periods (derived from the associated vector angular velocity under the assumption of solid-body rotation) first decrease precipitously, and then spin up gradually as the core secularly shrinks. This behavior is qualitatively similar for all our models, though we again caution that most of our other models have not ``settled down'' and asymptoted.

We emphasize that the induced spin-up we observe is not due to the spiral SASI $-$ we see the latter only for failed models \citep{Burrows2020}.  The spin up is due to the accretion of matter with an almost random range of impact parameters (``plus and minus''). The asymmetry of an explosion that arises from the spontaneous breaking of symmetry in 3D and that is correlated with whatever slight dipole asymmetry there may be in the blast can influence the post-explosion infall distribution, and provide it with a net direction.  The result is that the core is left with some spin, though it started in these calculations with none.

\begin{figure*}
    \includegraphics[width=\linewidth]{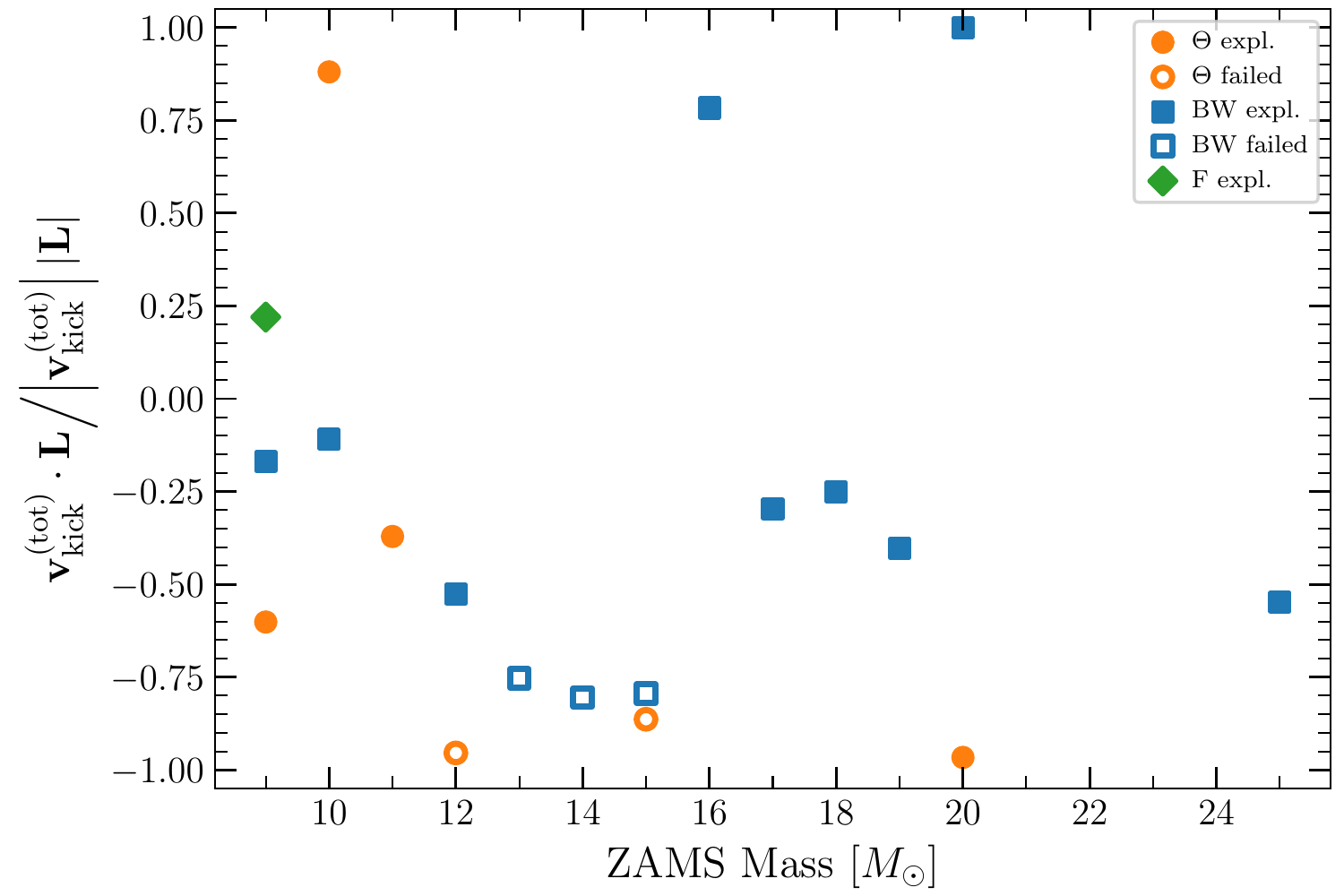}
    \caption{A plot demonstrating the degree of correlation between $\mathbf{v}_{\rm kick}^{(\rm tot)}$ and $\mathbf{L}_{\rm PNS}$ versus progenitor ZAMS mass. There seems to be a tentative tendency towards anti-alignment, but due to scatter and low number statistics this is at present only a tentative conclusion.}
    \label{fig:vtot_dot_L_zams}
\end{figure*}

Interestingly, we find that the direction of the kick/recoil is slightly anticorrelated with the direction of the final angular momentum.  Figure \ref{fig:vtot_dot_L_zams} depicts the cosine of the angle between the two. With some exceptions, for most models there is a clustering at negative values, though not $\cos(180^{\circ})$.   This is suggestive of the possible spin-kick correlation or alignment that has been inferred for radio pulsars \citep{Wang2006,Wang2007,Johnston2005,Johnston2007,Ng2007,Noutsos2013}.

Post-explosion anisotropic plume accretion onto the core will influence the dipole of the inner Y$_e$ distribution and neutrino optical depths in the radial direction \citep{Vartanyan2019b}.  The dipole of the neutrino emissions will reflect these quantities,
and, therefore, the neutrino kicks.  It is reasonable that this plume environment will also partially influence the spin up, so the loose correlation we see in Figure \ref{fig:vtot_dot_L_zams} is not that strange; it may suggest a relevance to the observed patterns for some radio pulsars.  However, this possibility needs to be much more rigorously checked.

\section{Conclusions}
\label{conclusions}

\begin{figure*}
    \includegraphics[width=\linewidth]{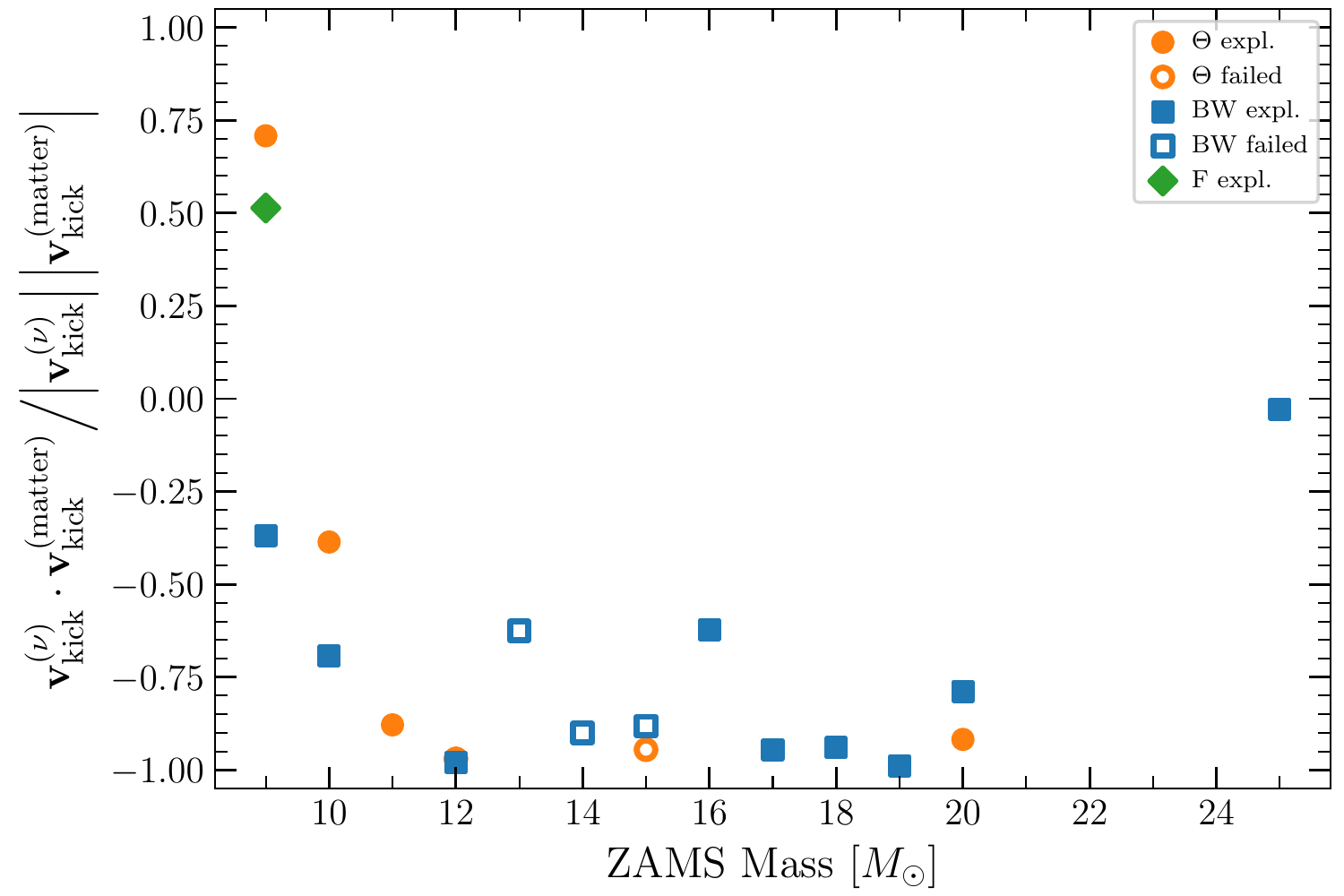}
    \caption{$\mathbf{v}_{\rm kick}^{(\nu)}\cdot\mathbf{v}_{\rm kick}^{(\rm matter)}/\left|\mathbf{v}_{\rm kick}^{(\nu)}\right| \left|\mathbf{v}_{\rm kick}^{(\rm matter)}\right|$, i.e. the cosine of the angle between $\mathbf{v}_{\rm kick}^{(\nu)}$ and $\mathbf{v}_{\rm kick}^{(\rm matter)}$, as a function of progenitor ZAMS mass for our various models. These two vectors are typically anti-aligned.}
    \label{fig:vnu_dot_vmat}
\end{figure*}

\begin{figure*}
    \includegraphics[width=\linewidth]{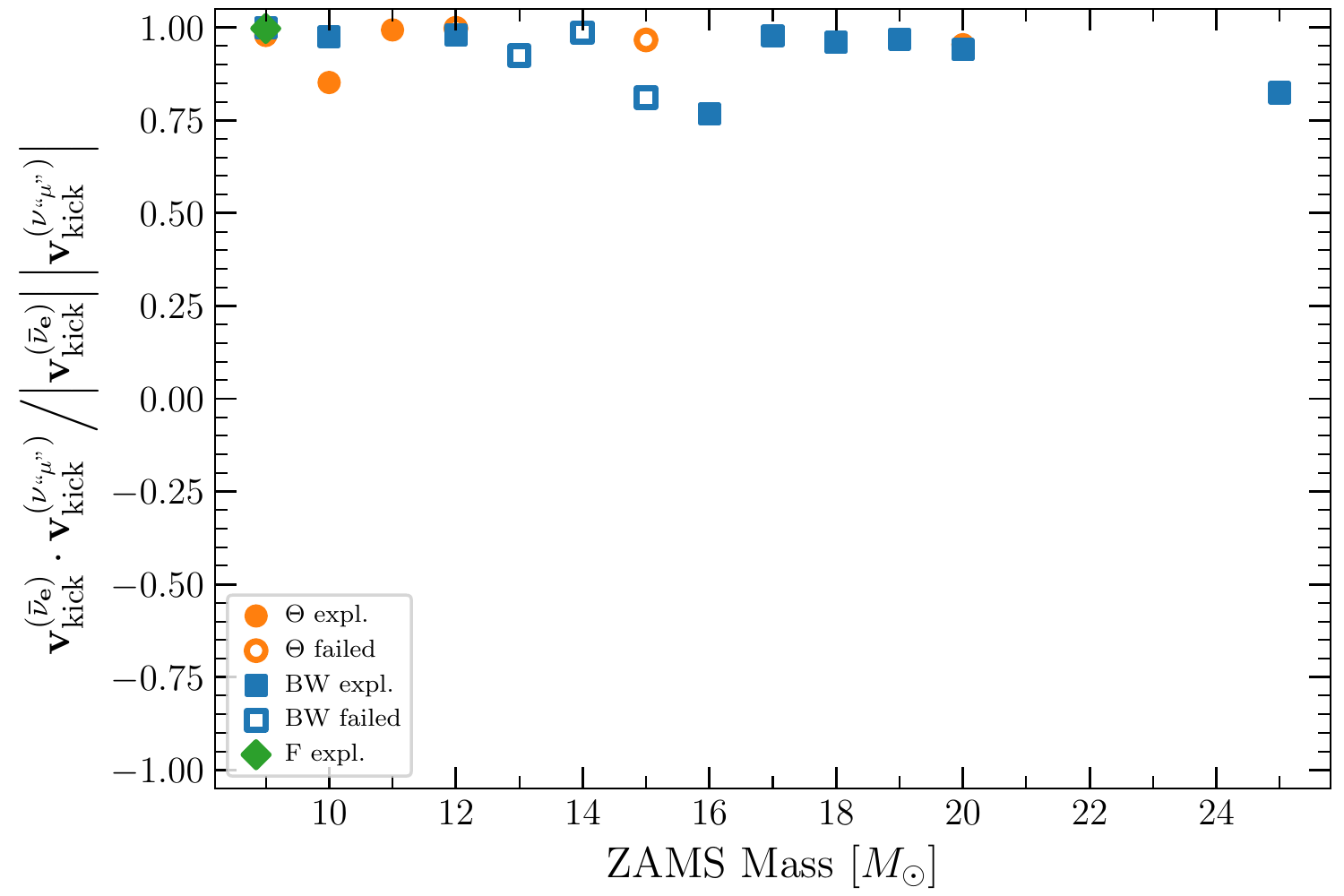}
    \caption{Same as Fig.~\ref{fig:vnu_dot_vmat}, but for $\mathbf{v}_{\rm kick}^{(\bar{\nu}_e)}$ and $\mathbf{v}_{\rm kick}^{(\nu_{\mu})}$. These vectors are well aligned in all our models.}
    \label{fig:vnu1_dot_vnu2}
\end{figure*}

\begin{figure*}
    \includegraphics[width=\linewidth]{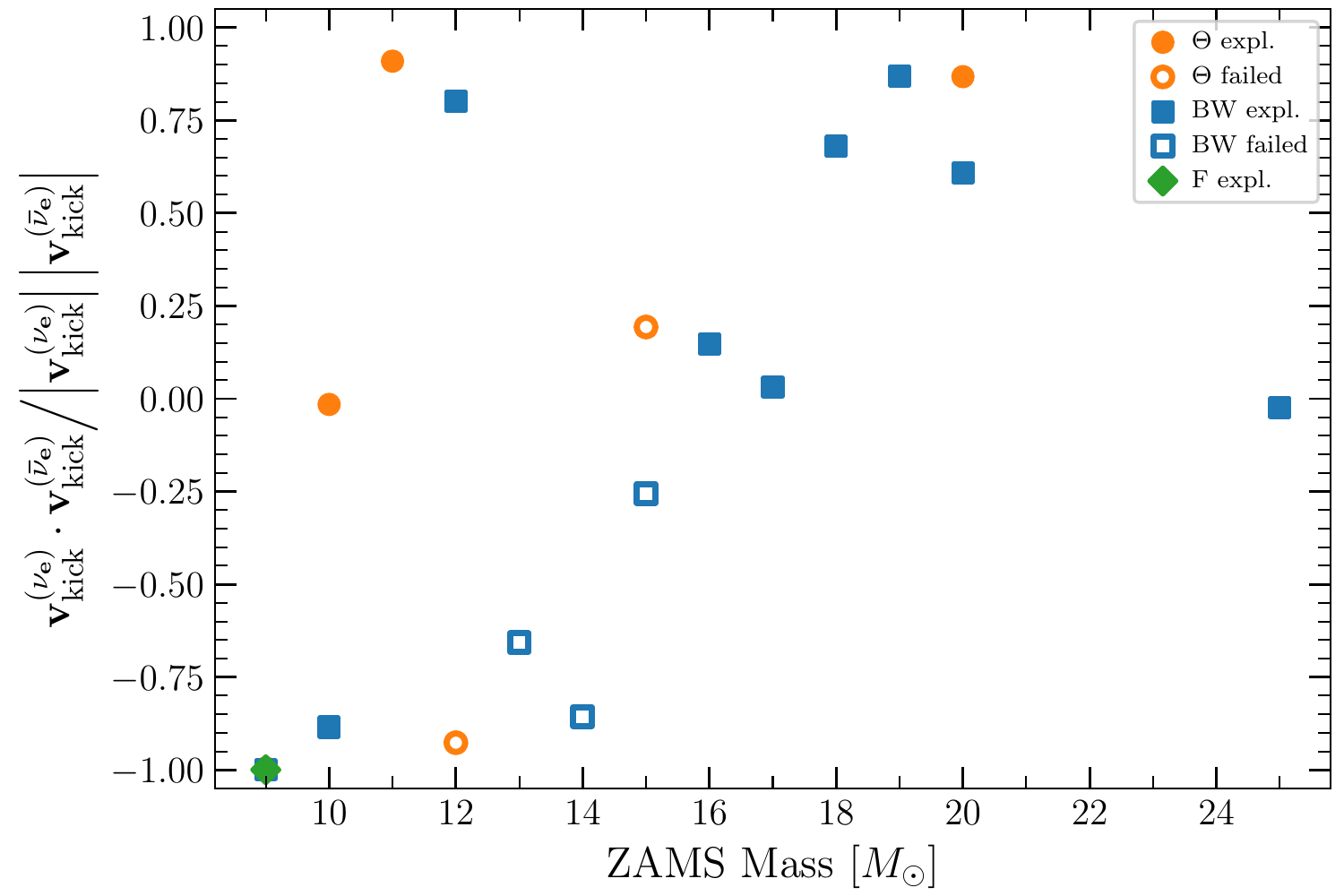}
    \caption{Same as Fig.~\ref{fig:vnu_dot_vmat}, but for $\mathbf{v}_{\rm kick}^{(\nu_e)}$ and $\mathbf{v}_{\rm kick}^{(\bar{\nu}_e)}$. There is no obvious systematic trend, though there is a slight tendency with increasing ZAMS mass towards alignment.}
    \label{fig:vnu0_dot_vnu1}
\end{figure*}

In this paper, for eighteen detailed 3D CCSN simulations, we have explored the kicks imparted to and the spins induced in the compact-object residues, be they neutron stars or stellar-mass black holes, of core collapse and supernova explosions.  We have found that the neutrino recoil due to anisotropic neutrino emissions can be an important factor affecting core recoil, comparable to and at times larger than the corresponding kick due to matter recoil.  This result, if true, would necessitate a revision of the general model of the origin of pulsar proper motions. Moreover, we find that the sign of the net neutrino momentum can be opposite to the sign of the corresponding matter recoil, though this is a statistical, not an absolute, statement. As a result, for a non-trivial fraction of CCSN explosions, the vector recoil and dipole asymmetry of the matter ejecta could be in the same direction; the standard model has them moving oppositely. However, in particular for the lowest mass progenitors (see Figure \ref{fig:vnu_dot_vmat}), the opposite might be true, so this is not an absolute conclusion.

From the only model (9.0 M$_{\odot}$ (F)) of the set that asymptotes in kick, spin, explosion energy, and PNS mass, we suggest that the duration of a significant dipole in the net neutrino emissions can be
of much shorter duration (perhaps $\sim$1 second) than the duration of the radiation of the neutron-star binding energy by the neutrinos (perhaps $\sim$tens of seconds).  This would allow a larger asymmetry (perhaps 5\%$-$10\%) to arise, but for a shorter time, to result in a product (Eqn. \ref{nkick}) that yields kicks in the observed pulsar range.

We find that the spin induced by the aspherical accretion of matter both just before and after the onset of explosion can leave the residue of collapse with a spin period that is comparable to those inferred for radio pulsars.  This is despite the fact that the initial models of this study were non-rotating. The agency of spin-up is the stochastic accretion of matter with different impact parameters.  However, there seems to be a slight anti-correlation between the direction of the induced spin and the net kick direction, both of which reflect in part the matter asymmetries in the explosion.  If true, this could explain the  inference of such a correlation among observed radio pulsars.

We find that the kicks imparted to black holes are due to the neutrino recoil alone.  While the magnitude of the imparted momentum can be comparable to that imparted to neutron stars, the likelihood that the final black hole mass is much larger than a neutron-star mass ($\sim$5$-$15 M$_{\odot}$) results in kick speeds that are $\le$100 km s$^{-1}$ most of the time.

It should be remembered that most of the 3D models in this study have not achieved their final states and that the interesting observables, such as explosion energy and recoil kick, have not converged. Simulations to many seconds after bounce are required for all but the least massive progenitors in order to reach that state \citep{Muller2017,Burrows2021}. Therefore, one should view our conclusions and inferred correlations as interesting and suggestive, but merely provisional.  Nevertheless, what has emerged from this investigation is perhaps a tantalizing glimpse of the emerging opportunities to couple modern supernova theory with pulsar and black-hole observations more tightly than has been possible in the past.

\section*{Acknowledgements}

We thank David Vartanyan, Tianshu Wang, Chris White, David Radice, and Hiroki Nagakura for fruitful discussions and/or past collaborations. We acknowledge support from the U.~S.\ Department of Energy Office of Science and the Office of Advanced Scientific Computing Research via the Scientific Discovery through Advanced Computing (SciDAC4) program and Grant DE-SC0018297 (subaward 00009650) and support from the U.~S.\ National Science Foundation (NSF) under Grants AST-1714267 and PHY-1804048 (the latter via the Max-Planck/Princeton Center (MPPC) for Plasma Physics). The bulk of the computations presented in this work were performed on Blue Waters under the sustained-petascale computing project, which was supported by the National Science Foundation (awards OCI-0725070 and ACI-1238993) and the state of Illinois. Blue Waters was a joint effort of the University of Illinois at Urbana--Champaign and its National Center for Supercomputing Applications. The two $9\ M_{\sun}$ models were simulated on the Frontera cluster (under awards AST20020 and AST21003), and this research is part of the Frontera computing project at the Texas Advanced Computing Center \citep{Stanzione2020}. Frontera is made possible by NSF award OAC-1818253. Additionally, a generous award of computer time was provided by the INCITE program, enabling this research to use resources of the Argonne Leadership Computing Facility, a DOE Office of Science User Facility supported under Contract DE-AC02-06CH11357. Finally, the authors acknowledge computational resources provided by the high-performance computer center at Princeton University, which is jointly supported by the Princeton Institute for Computational Science and Engineering (PICSciE) and the Princeton University Office of Information Technology, and our continuing allocation at the National Energy Research Scientific Computing Center (NERSC), which is supported by the Office of Science of the U.~S.\ Department of Energy under contract DE-AC03-76SF00098.

\section*{Data Availability}

The data underlying this article will be shared on reasonable request to the corresponding author.



\bibliographystyle{mnras}
\bibliography{citations}



\begin{figure*}
    \includegraphics[width=1\linewidth]{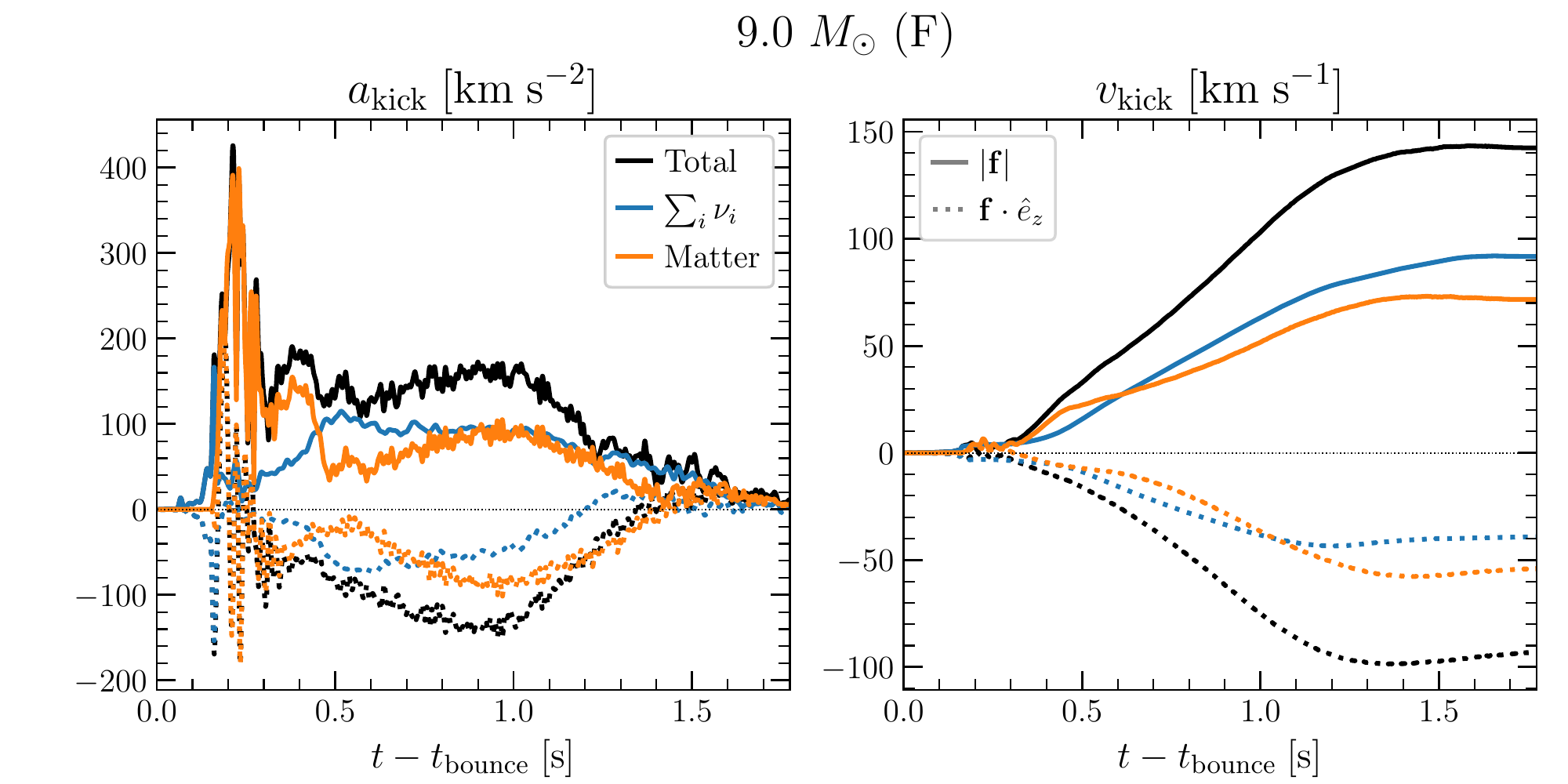}
    \caption{PNS kick accelerations (left) and time-integrated velocities (right) as a function of time after bounce for the 9.0 M$_{\odot}$ (F) model.
    In order to reduce the jagged nature of the accelerations, these data have been smoothed/convolved (for visual purposes only) with a Hanning window of width 21 ms. The total kick (black) has also been decomposed into the contributions from neutrinos (blue) and matter (orange).
    Solid lines show the vector magnitude while dotted lines show the $z$ component of the corresponding vector.
    For this 9.0 M$_{\odot}$ model, the matter and neutrino components of the kick are approximately aligned, with the asymptotic matter kick being roughly $\sim$80\% the magnitude of the neutrino kick.}
    \label{fig:9.0_kick_evo}
\end{figure*}

\begin{figure*}
    \includegraphics[width=1\linewidth]{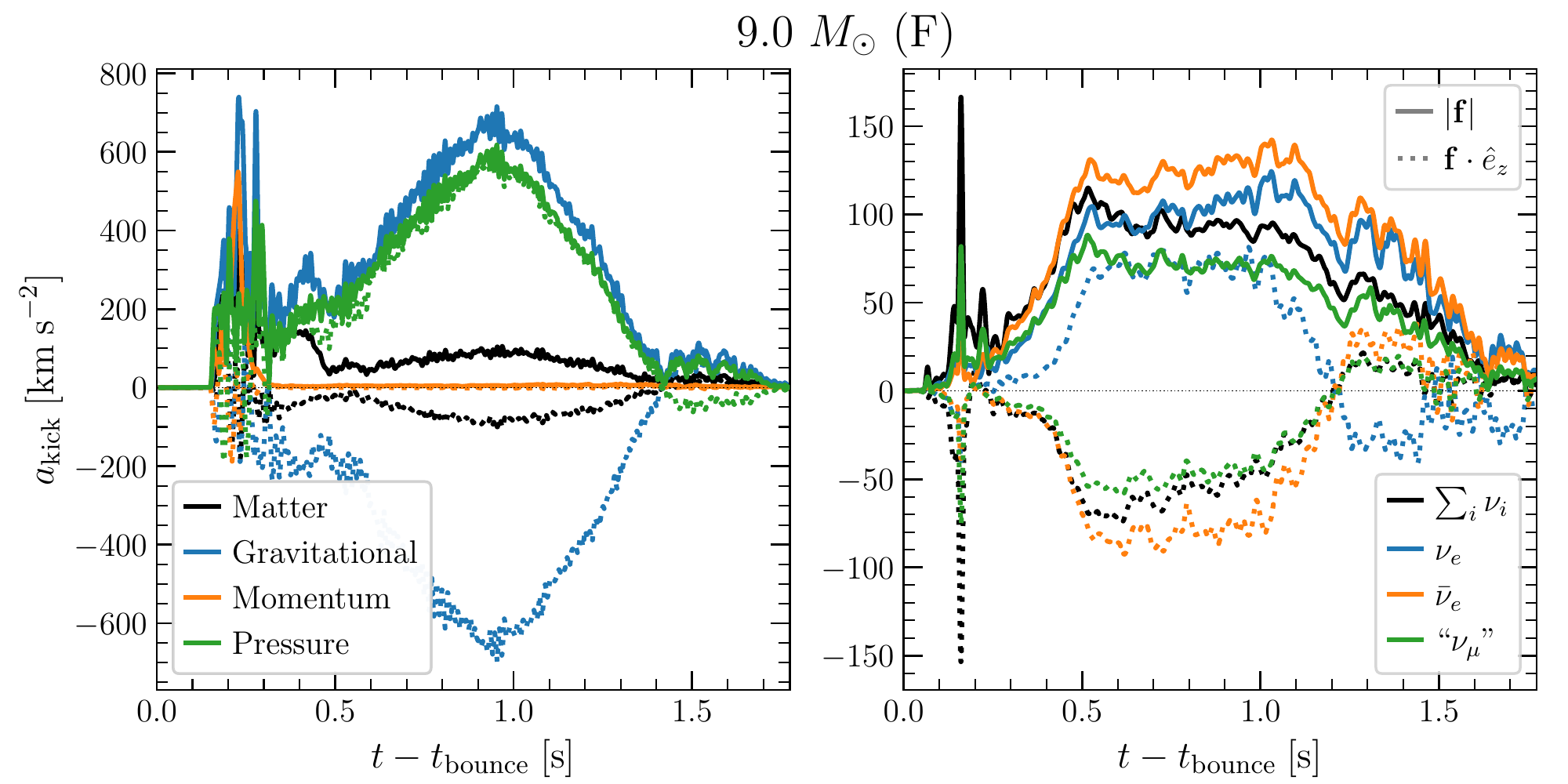}
    \caption{Further decomposition of PNS kick accelerations as a function of time after bounce for the 9.0 M$_{\odot}$ progenitor.
    These data have been smoothed/convolved (for visual purposes only) with a Hanning window of width 21 ms. The total matter contribution of the kick-acceleration (left, black) is accompanied by the magnitudes of the corresponding gravitational (blue), momentum (orange), and pressure (green) components (see \S\ref{sec:kicks}).
    The vector summed acceleration of all neutrino species (right, black) is compared with the magnitudes of the contributions of the three separate neutrino species. As in Fig. \ref{fig:9.0_kick_evo}, solid lines show the vector magnitude, while dotted lines show the $z$ component of the corresponding vectors, demonstrating that the $\nu_e$ and $\bar{\nu}_e$ components are nearly anti-aligned for this model.
    }
    \label{fig:9.0_detailed_kick_evo}
\end{figure*}

\begin{figure*}
    \includegraphics[width=1\linewidth]{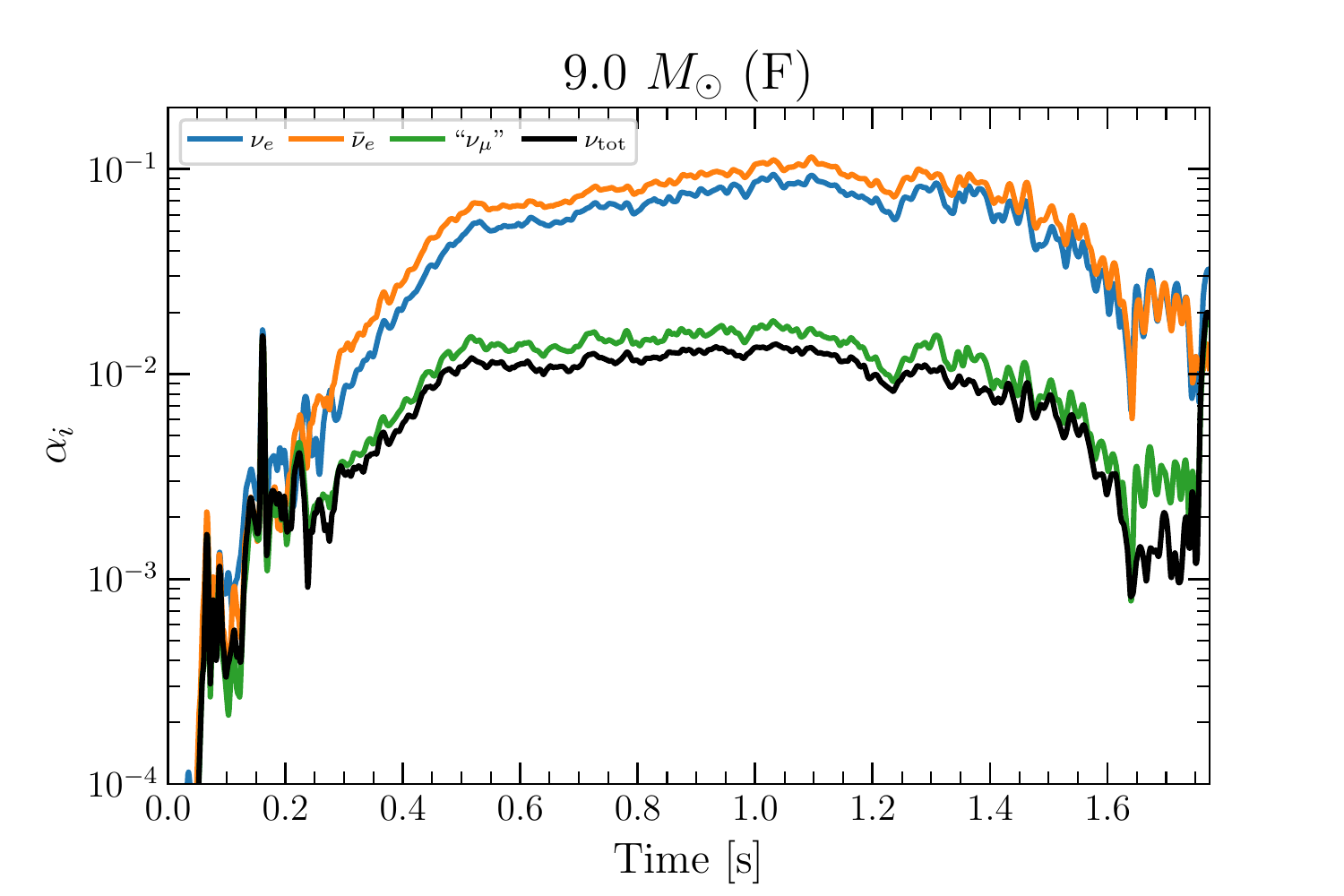}
    \caption{Instantaneous anisotropy parameters (see Eqn.~\ref{eqn:aniso1}) associated with three neutrino species, and their vector summed effect (see legend), as a function of time for the 9.0 $M_{\sun}$ model. Note that the anisotropy associated with $\nu_e$ is nearly anti-aligned with that of the $\bar{\nu}_e$ (see Fig.~\ref{fig:9.0_detailed_kick_evo}), causing the black line (representing the net neutrino effect) to be comparable to but lower than the green line representing the ``$\nu_{\mu}$''s.
    In order to reduce the jagged nature of $\alpha_i$, these data have been smoothed/convolved (for visual purposes only) with a Hanning window of width 21 ms.}
    \label{fig:9.0_anisotropy}
\end{figure*}

\begin{figure*}
    \includegraphics[width=1\linewidth]{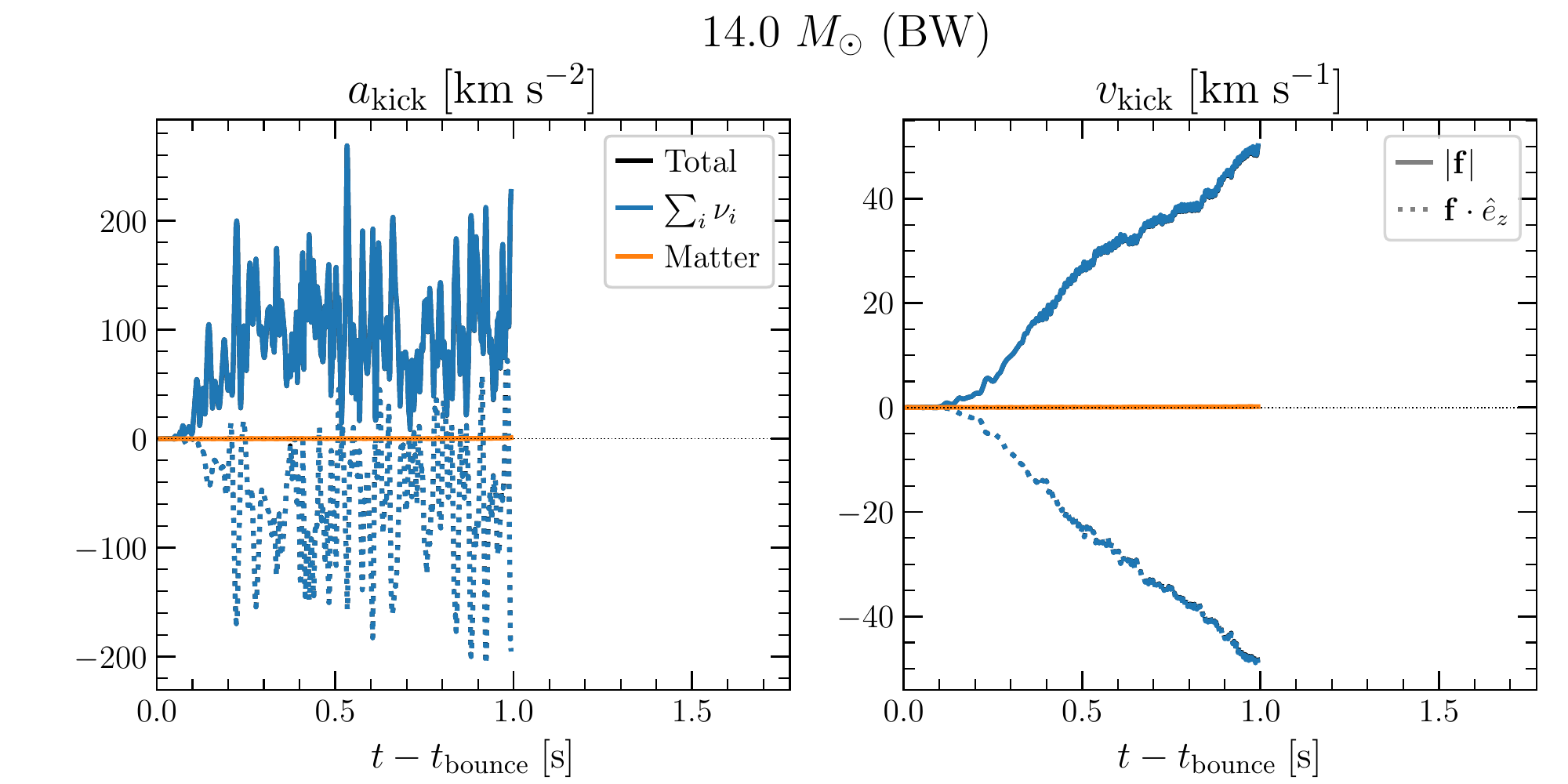}
    \caption{Same as Fig.~\ref{fig:9.0_kick_evo}, but for the 14.0 $M_{\sun}$ model from \citet{Burrows2020}, which fails to explode. In this case the kick contribution from matter is negligible.
    }
    \label{fig:14.0-oa_kick_evo}
\end{figure*}

\begin{figure*}
    \begin{center}
    \includegraphics[width=0.5\linewidth]{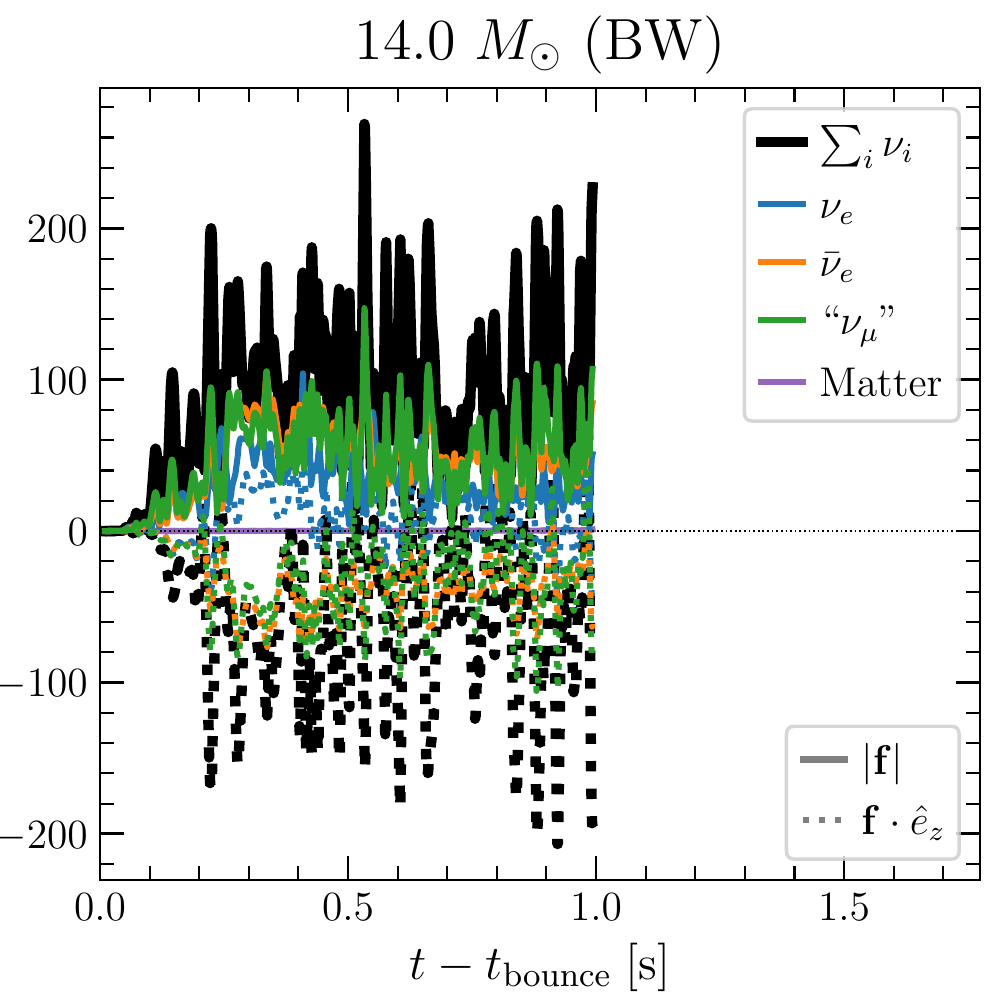}
    \end{center}
    \caption{
    Similar to the right panel of Fig.~\ref{fig:9.0_detailed_kick_evo}, but includes the matter contribution (purple), and for the 14.0 $M_{\sun}$ (BW) model from \citet{Burrows2020}. Note that the matter contribution (light purple curve) hovers near zero acceleration for this non-exploding model.
    }
    \label{fig:14.0-oa_detailed_kick_evo}
\end{figure*}

\begin{figure*}
    \includegraphics[width=1\linewidth]{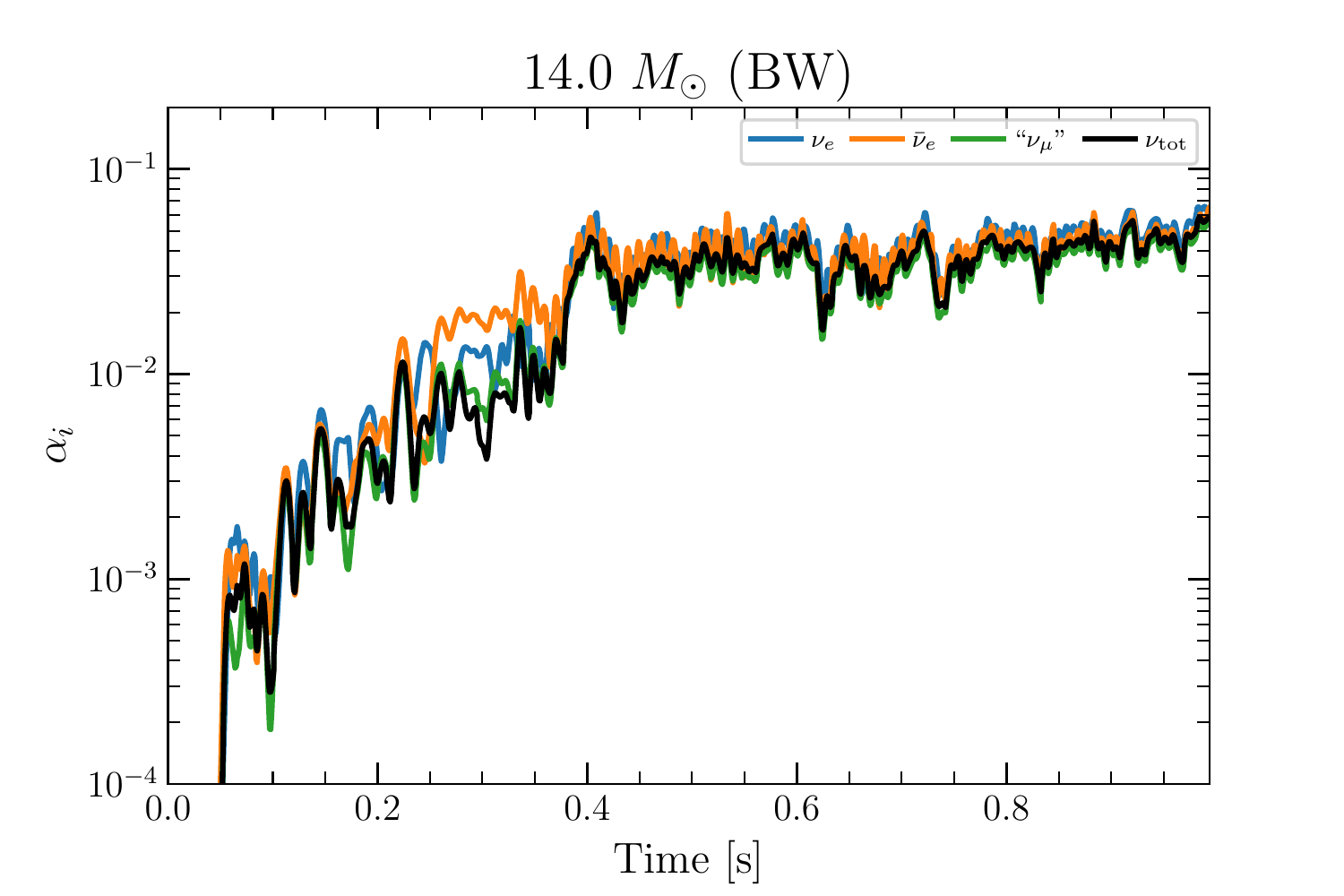}
    \caption{Same as Fig.~\ref{fig:9.0_anisotropy}, but for the 14.0 $M_{\sun}$ (BW) model from \citet{Burrows2020}. Similar to the 9.0 $M_{\sun}$ (F) model, the anisotopies associated with $\nu_e$ and $\bar{\nu}_e$ nearly cancel, making the net anisotropy closely track that of ``$\nu_{\mu}$s," which for this model is comparable in magnitude to the other anisotropy components.}
    \label{fig:14.0-oa_anisotropy}
\end{figure*}


\appendix
\section{Additional Figures}
\begin{figure*}
    \includegraphics[width=1\linewidth]{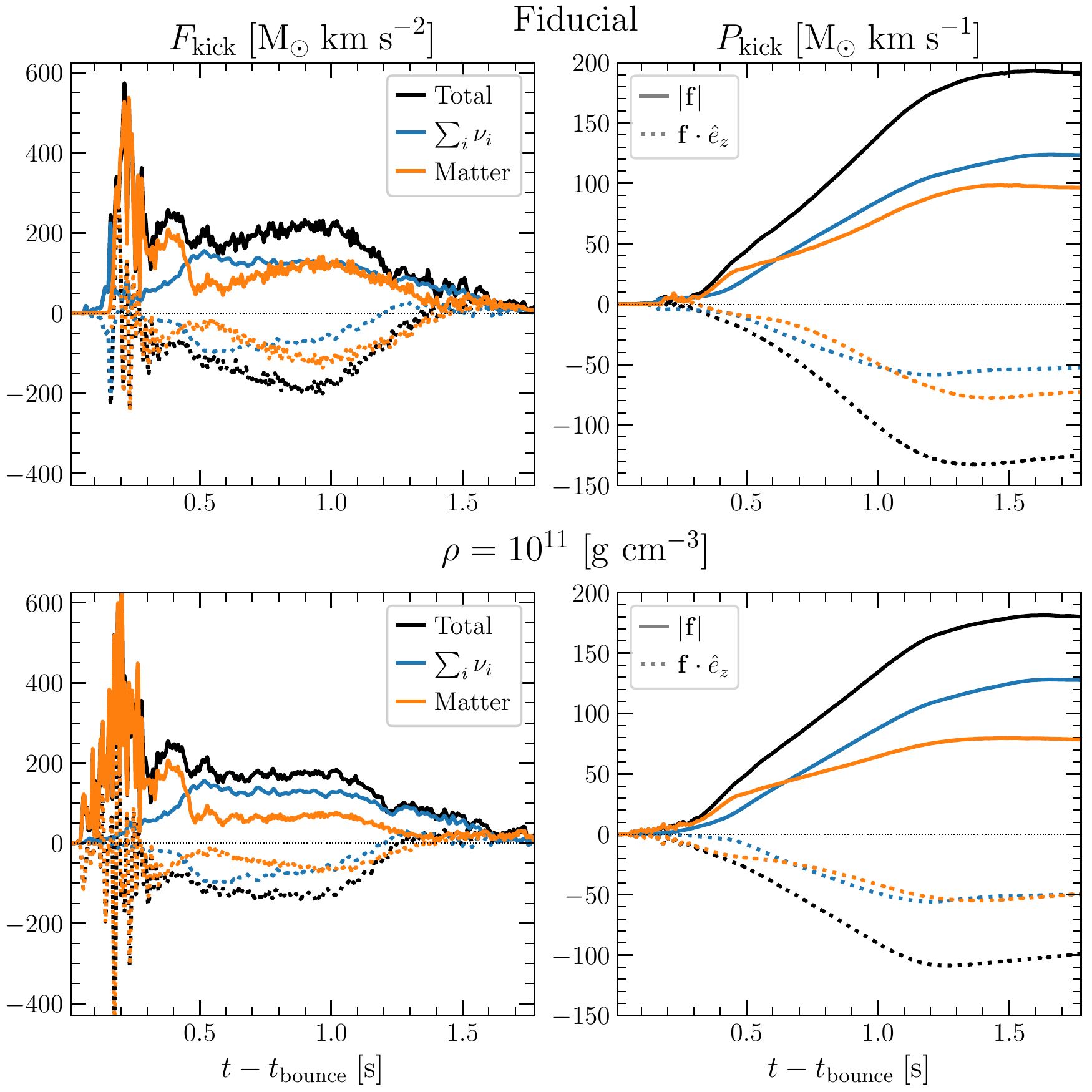}
    \caption{
    Similar to Fig.~\ref{fig:9.0_kick_evo}, but showing the kick force (left) and momentum (right) as opposed to acceleration and velocity, respectively, for the same 9.0 M$_\odot$ (F) model.
    The top integrates the momentum forces over our fiducial integration surface (see the end of Section~\ref{sec:kicks}), while the bottom row utilises the isodensity surface of $\rho=10^{11}$ g cm$^{-3}$. The differences are within $\sim$10\%, but we believe our procedure produces the more physical and complete asymtotic result.
    }
    \label{fig:kick_methods}
\end{figure*}

\begin{figure*}
    \includegraphics[width=1\linewidth]{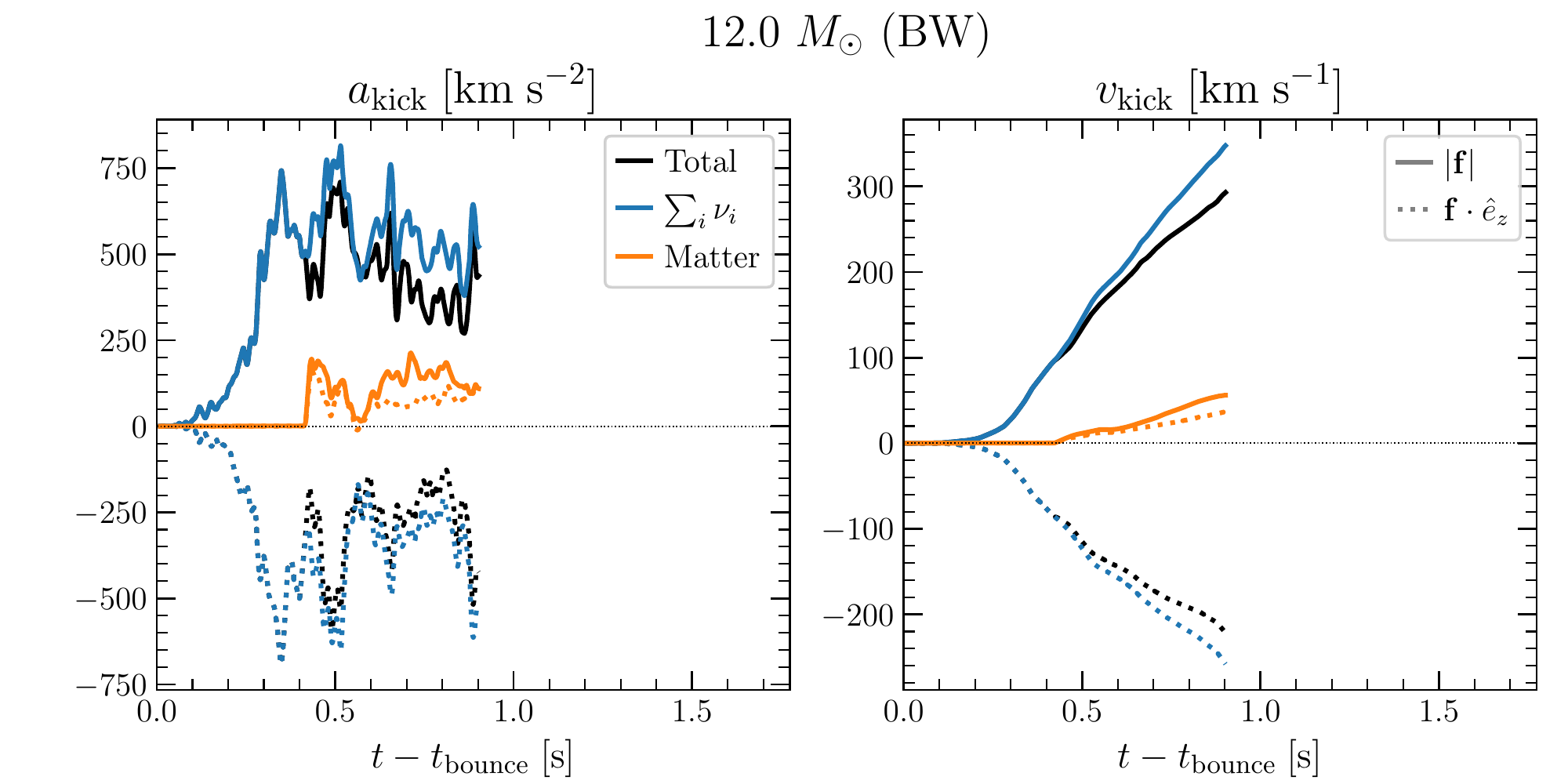}
    \caption{Same as Fig.~\ref{fig:9.0_kick_evo}, but for the 12.0 $M_{\sun}$ (BW) model from \citet{Burrows2020}.
    }
    \label{fig:12.0-oa_kick_evo}
\end{figure*}

\begin{figure*}
    \includegraphics[width=1\linewidth]{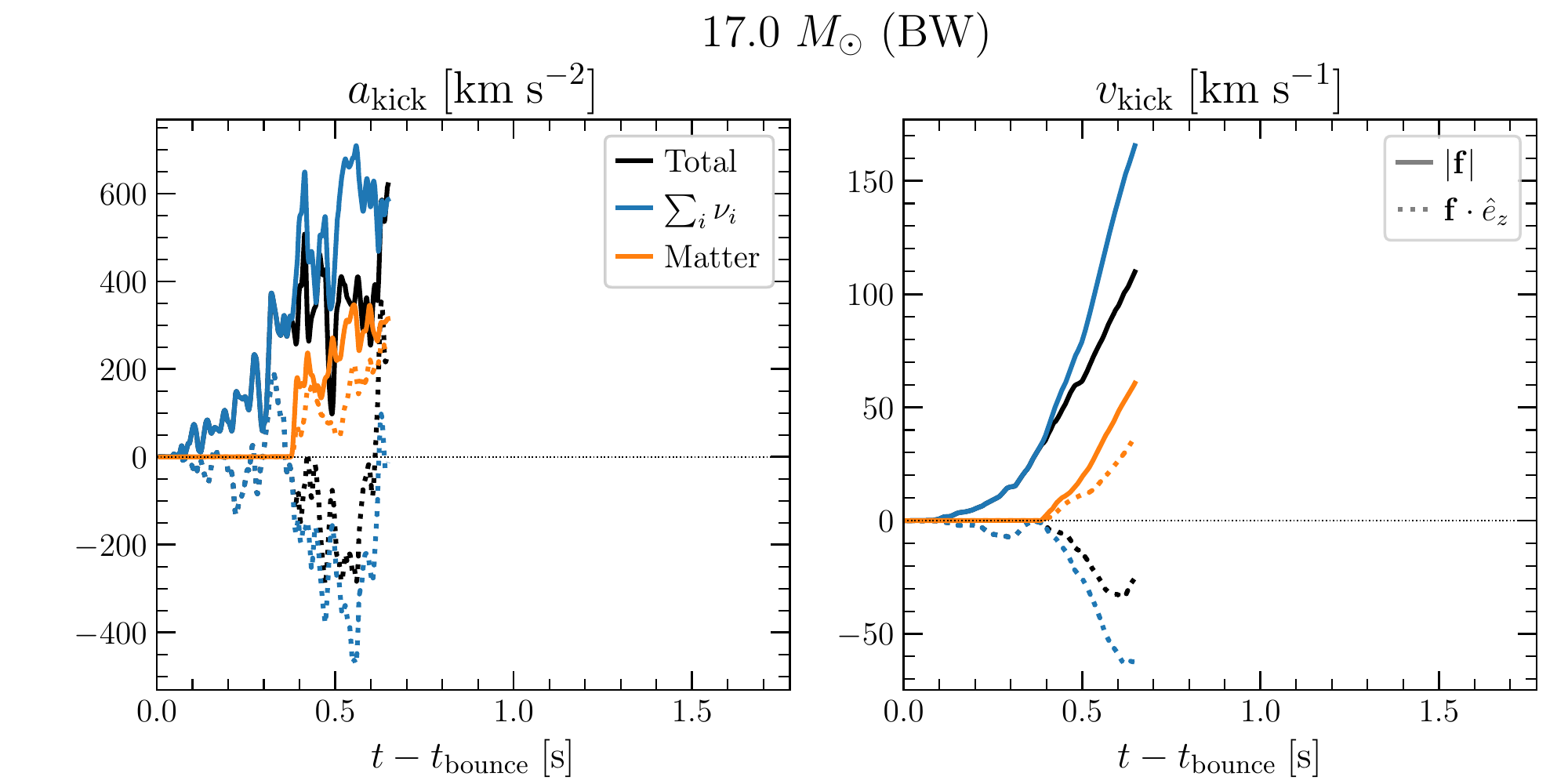}
    \caption{Same as Fig.~\ref{fig:9.0_kick_evo}, but for the 17.0 $M_{\sun}$ (BW) model from \citet{Burrows2020}.
    }
    \label{fig:17.0-oa_kick_evo}
\end{figure*}

\begin{figure}
    \includegraphics[width=1\linewidth]{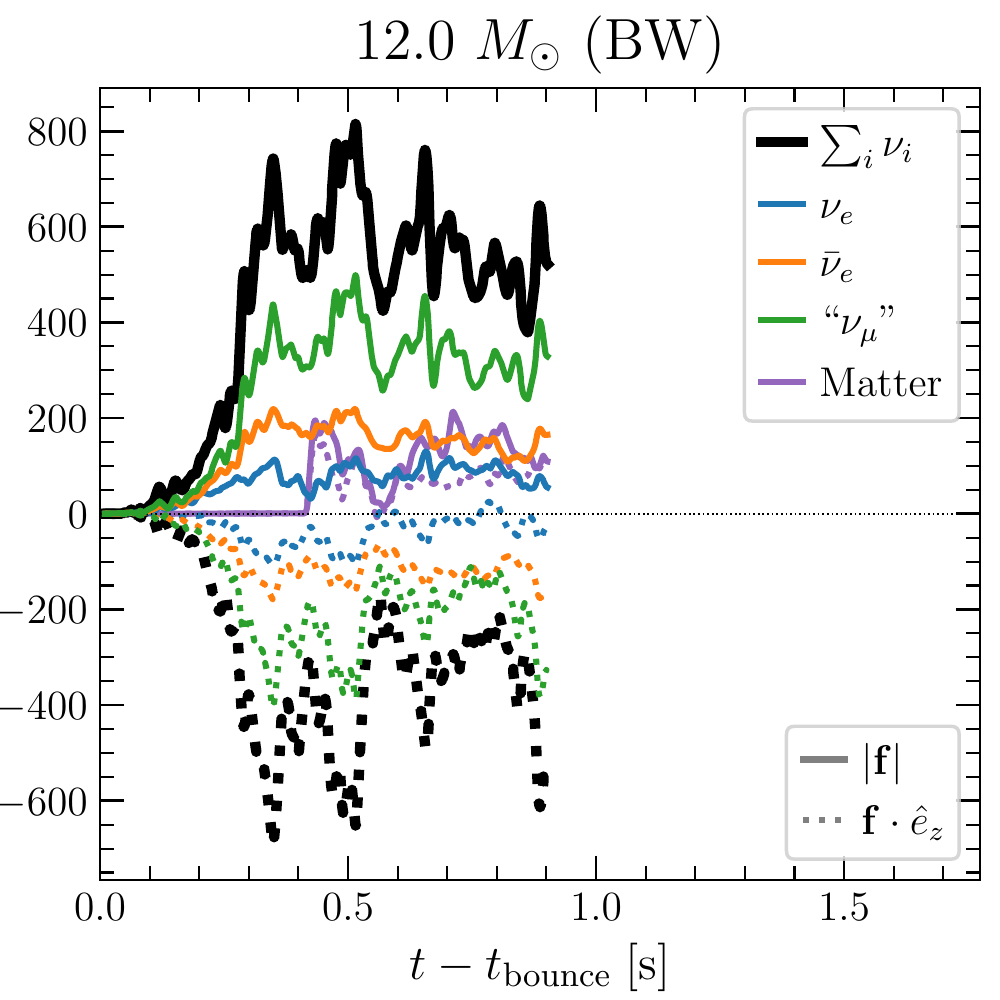}
    \caption{Similar to the right panel of Fig.~\ref{fig:9.0_detailed_kick_evo}, but includes the matter contribution (purple), and for the 12.0 $M_{\sun}$ (BW) model from \citet{Burrows2020}.
    }
    \label{fig:12.0-oa_detailed_kick_evo}
\end{figure}

\begin{figure}
    \includegraphics[width=1\linewidth]{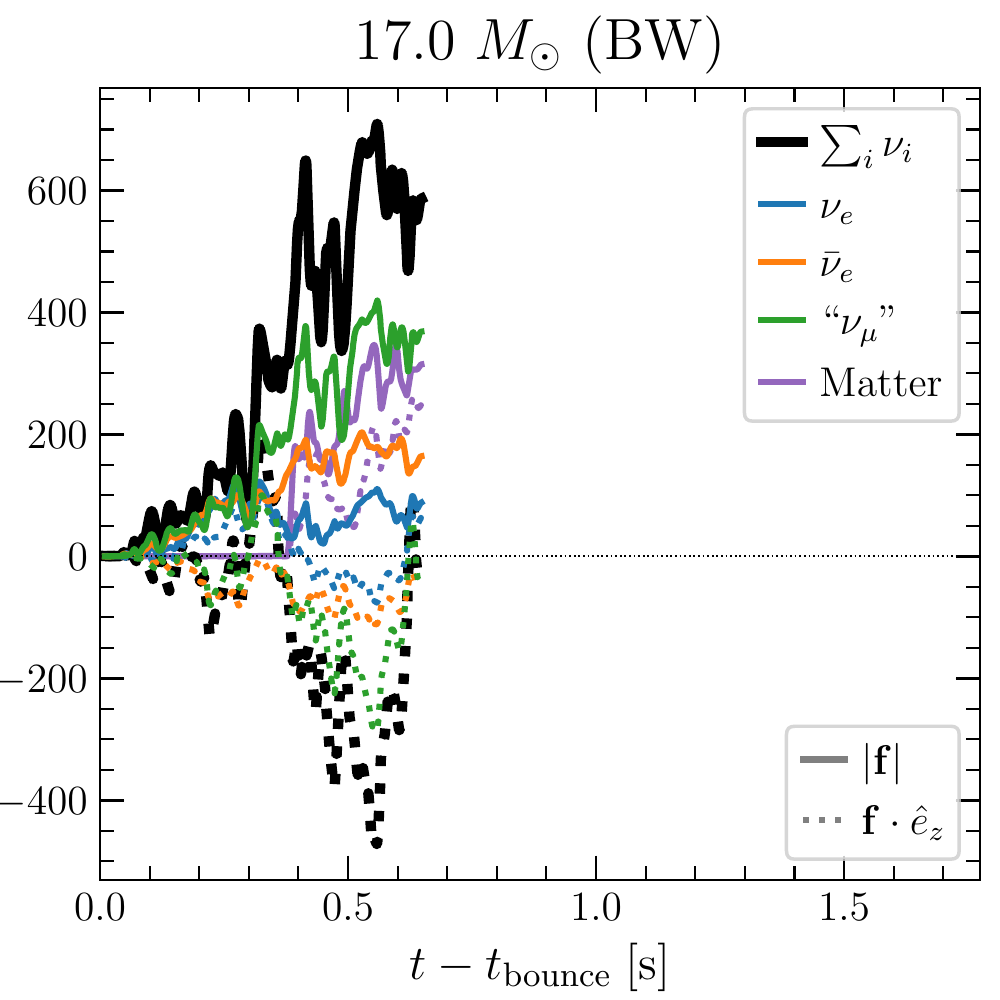}
    \caption{Same as Fig.~\ref{fig:12.0-oa_detailed_kick_evo}, but for the 17.0 $M_{\sun}$ (BW) model from \citet{Burrows2020}.
    }
    \label{fig:17.0-oa_detailed_kick_evo}
\end{figure}

\begin{figure}
    \includegraphics[width=1\linewidth]{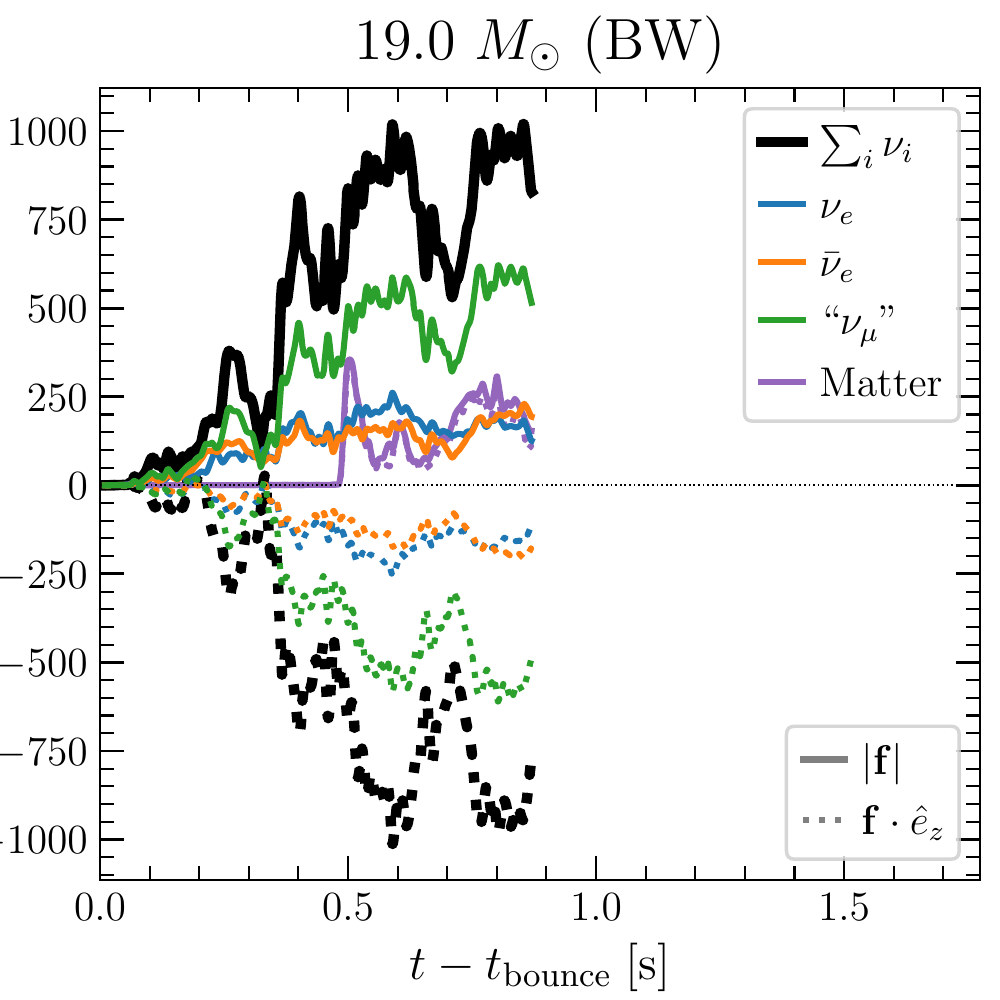}
    \caption{Same as Fig.~\ref{fig:12.0-oa_detailed_kick_evo}, but for the 19.0 $M_{\sun}$ (BW) model from \citet{Burrows2020}.
    }
    \label{fig:19.0-oa_detailed_kick_evo}
\end{figure}

\begin{figure*}
    \includegraphics[width=1\linewidth]{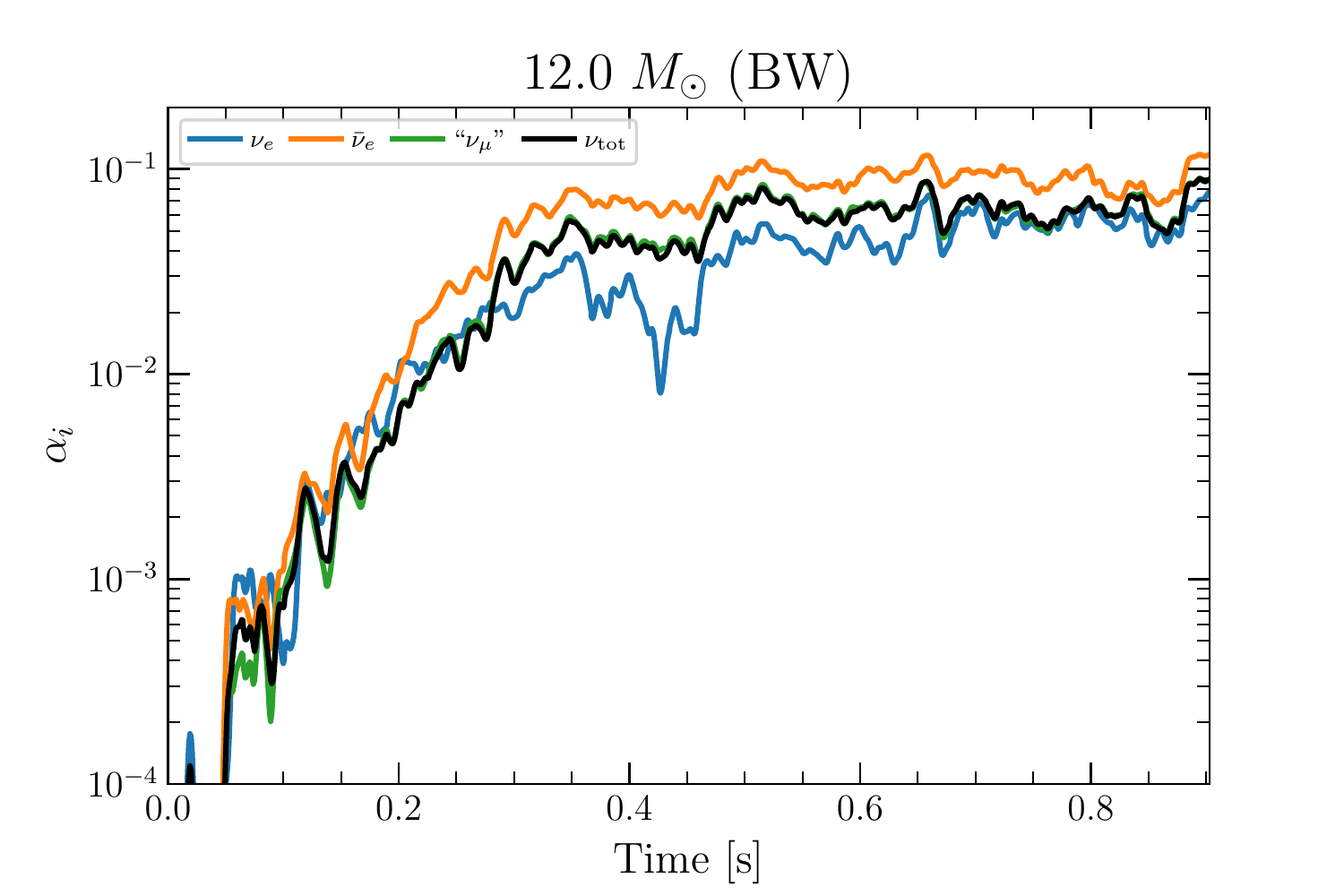}
    \caption{Same as Fig.~\ref{fig:9.0_anisotropy}, but for the 12.0 $M_{\sun}$ (BW) model from \citet{Burrows2020}. Similar to the 9.0 $M_{\sun}$ model, the anisotopy associated with $\nu_e$ and $\bar{\nu}_e$ nearly cancel, making the net anisotropy closely track that of the $\nu_{\mu}$s.}
    \label{fig:12.0-oa_anisotropy}
\end{figure*}

\begin{figure*}
    \includegraphics[width=1\linewidth]{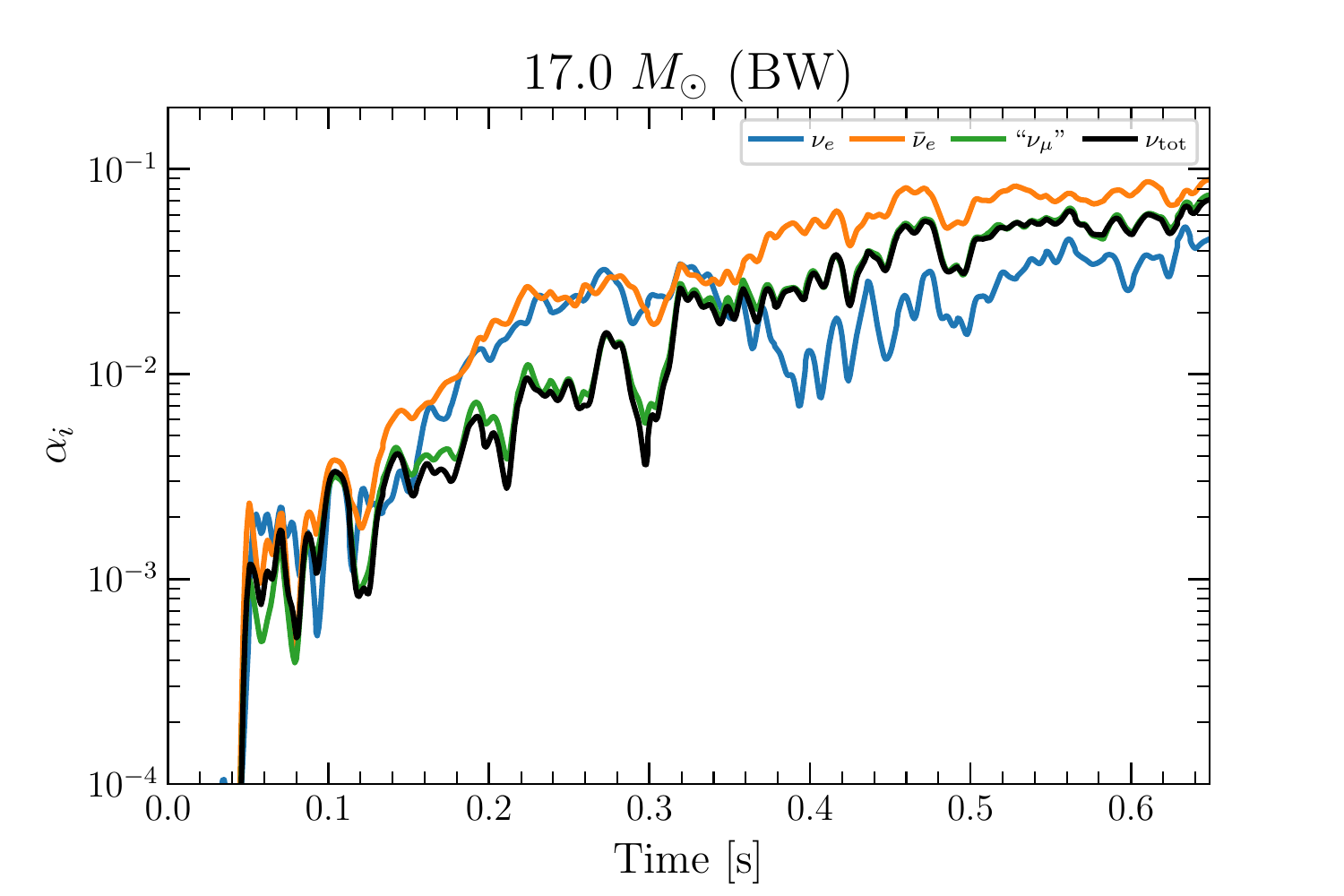}
    \caption{Same as Fig.~\ref{fig:9.0_anisotropy}, but for the 17.0 $M_{\sun}$ (BW) model from \citet{Burrows2020}.
    }
    \label{fig:17.0-oa_anisotropy}
\end{figure*}

\begin{figure*}
    \includegraphics[width=1\linewidth]{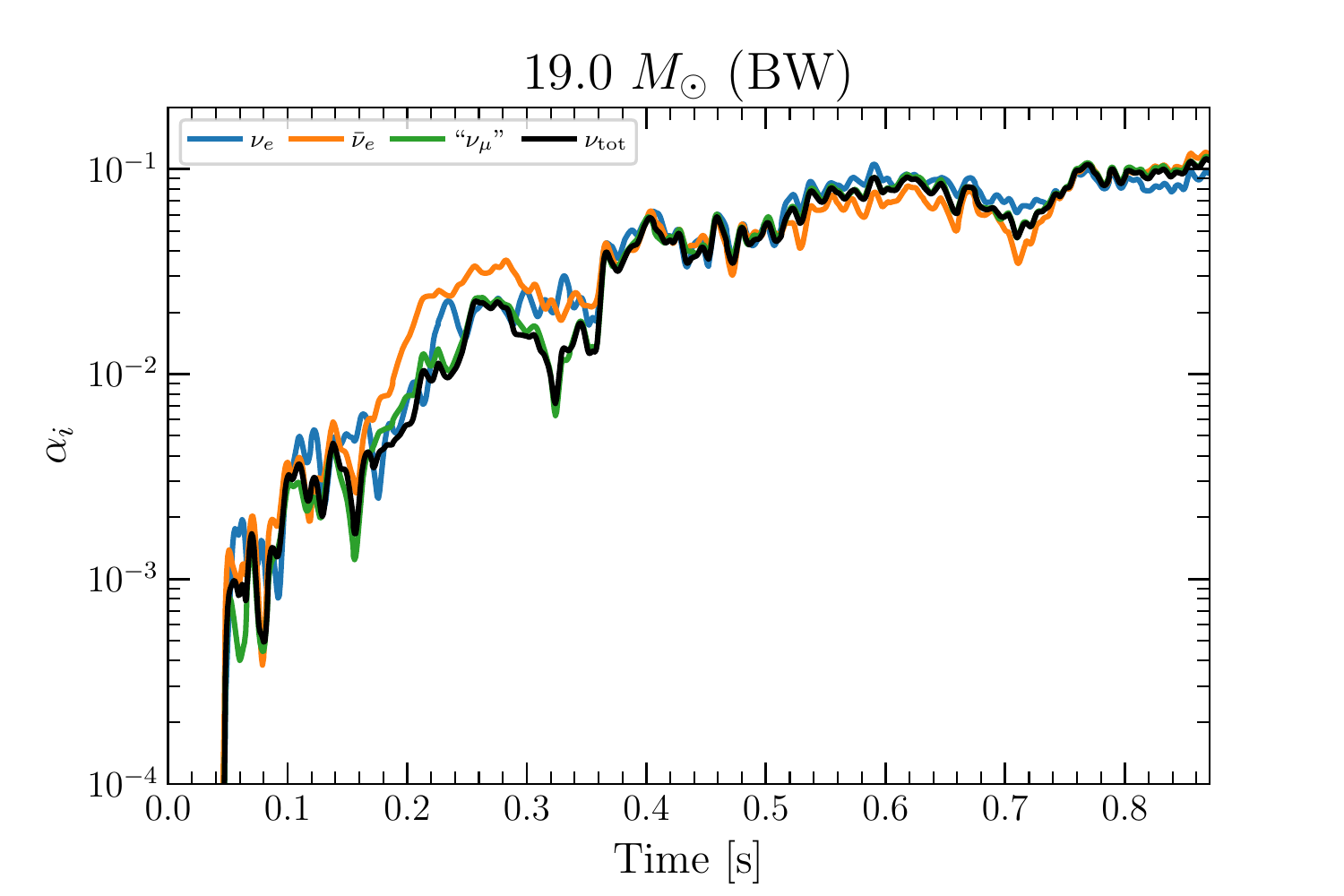}
    \caption{Same as Fig.~\ref{fig:9.0_anisotropy}, but for the 19.0 $M_{\sun}$ (BW) model from \citet{Burrows2020}. In this model, despite the anisotropy of the $\nu_e$ and $\bar{\nu}_e$ being aligned, the net anisotropy closely follows that of ``$\nu_{\mu}$''. This is because ``$\nu_{\mu}$'' dominates the neutrino flux (and kick) as it represents the sum effect of what are in nature four distinct neutrino species.}
    \label{fig:19.0-oa_anisotropy}
\end{figure*}

This appendix contains additional figures. 
We show two different choices of integration surfaces for the kicks of the 9.0 M$_\odot$ in Figure~\ref{fig:kick_methods}.
The remaining figures in this appendix are
for the 12.0 $M_{\sun}$ (BW), 17.0 $M_{\sun}$ (BW), and 19.0 $M_{\sun}$ (BW) models from \citet{Burrows2020}.
Figures \ref{fig:12.0-oa_kick_evo}, \ref{fig:17.0-oa_kick_evo} and \ref{fig:19.0-oa_detailed_kick_evo} are the same as Fig.~\ref{fig:9.0_kick_evo} except for the 12, 17, and 19 $M_{\sun}$ models respectively.
Figures \ref{fig:12.0-oa_detailed_kick_evo}, \ref{fig:17.0-oa_detailed_kick_evo}, and \ref{fig:19.0-oa_detailed_kick_evo} are the same as Fig.~\ref{fig:9.0_detailed_kick_evo} except for the 12, 17, and 19 $M_{\sun}$ models respectively.
Figures \ref{fig:12.0-oa_anisotropy}, \ref{fig:17.0-oa_anisotropy}, and \ref{fig:19.0-oa_anisotropy} are the same as Fig.~\ref{fig:9.0_anisotropy} except for the 12, 17, and 19 $M_{\sun}$ models respectively.


\begin{figure*}
    \includegraphics[width=1\linewidth]{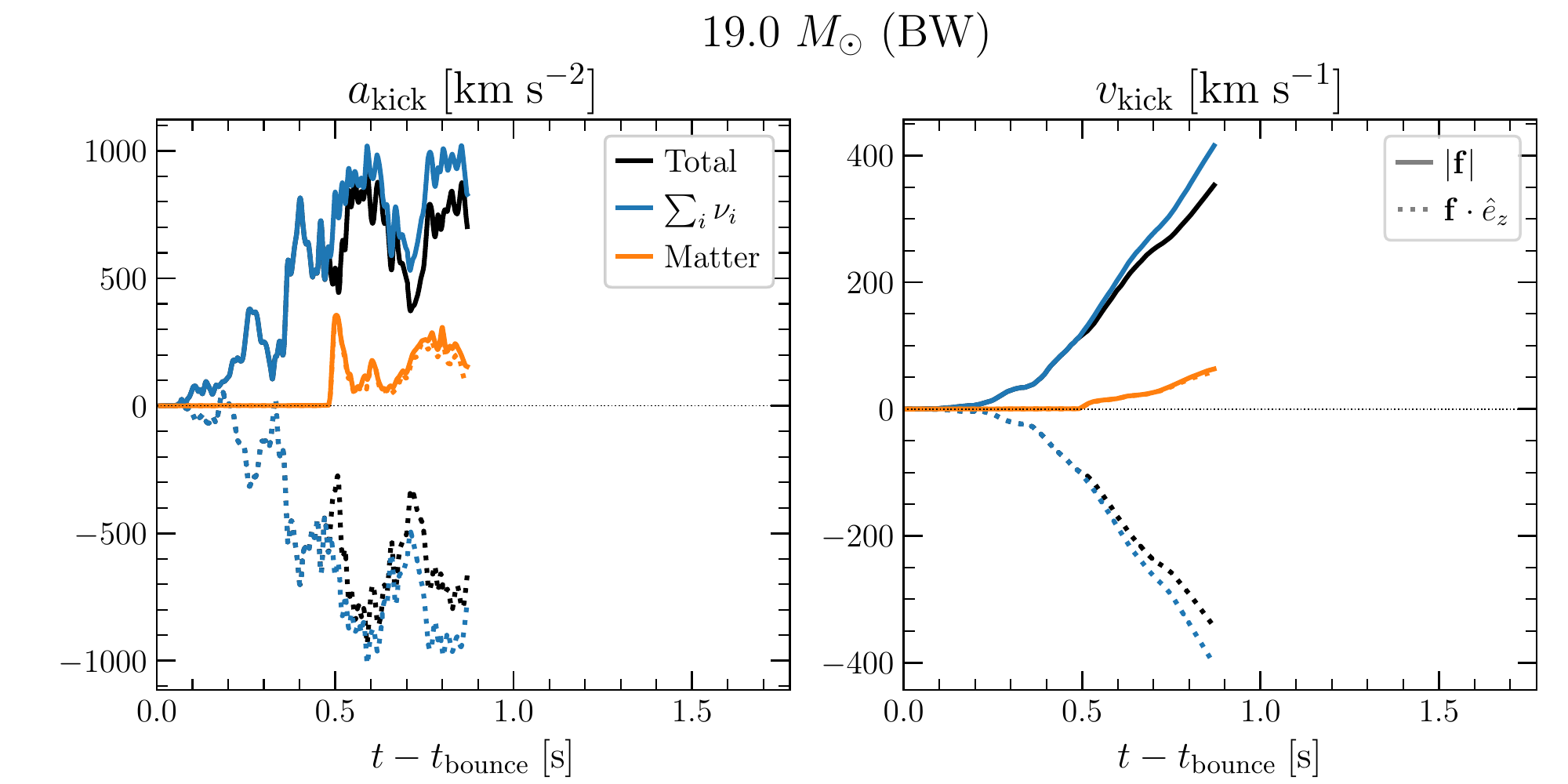}
    \caption{Same as Fig.~\ref{fig:9.0_kick_evo}, but for the 19.0 $M_{\sun}$ model from \citet{Burrows2020}. For this model the matter and neutrino components are roughly anti-aligned with the neutrino component dominating by a factor of several.
    }
    \label{fig:19.0-oa_kick_evo}
\end{figure*}


\bsp	
\label{lastpage}
\end{document}